\documentclass[epjST]{svjour}
\usepackage{graphicx}
\usepackage[caption=false]{subfig}
\usepackage{amssymb}
\usepackage{esint}
\begin{document}
\title{On the Difference between FOPT and CIPT for Hadronic Tau Decays}
\author{Andr\'e H. Hoang\inst{1,2}\fnmsep\thanks{\email{andre.hoang@univie.ac.at} } \and 
	Christoph Regner\inst{1}\fnmsep\thanks{\email{christoph.regner@univie.ac.at}\hfill UWThPh 2021-5}}
\institute{University of Vienna, Faculty of Physics, Boltzmanngasse 5, A-1090 Wien, Austria \and 
	       Erwin Schr\"odinger International Institute for Mathematics and Physics,
	       University of Vienna, Boltzmanngasse 9, A-1090 Wien, Austria }
\abstract{
In this article we review the results of our recent work on the difference between the Borel representations of $\tau$ hadronic spectral function moments obtained with the CIPT and FOPT methods. For the presentation of the theoretical results we focus on the large-$\beta_0$ approximation, where all expressions can be written down in closed form, and we comment on the generalization to full QCD. The results may explain the discrepancy in the behavior of the FOPT and CIPT series that has been the topic of intense discussions in previous literature and which represents a major part of the theoretical uncertainties in current strong coupling determinations from hadronic $\tau$ decays. The findings also imply that the OPE corrections for FOPT and CIPT differ and that the OPE corrections for CIPT do not have standard form.
} %end of abstract
\maketitle
\section{Introduction}
\label{sec:introduction}

Moments of the $\tau$ hadronic spectral functions obtained from LEP~\cite{Davier:2013sfa,Ackerstaff:1998yj} constitute a major tool for precise determinations of the strong coupling $\alpha_s$. The theoretical predictions for the spectral function moments in the massless quark limit are based on the vacuum polarization function
$\Pi(p^2)$, which is known perturbatively to 5 loops (i.e.\ ${\cal O}(\alpha_s^4)$)~\cite{Gorishnii:1990vf,Surguladze:1990tg,Baikov:2008jh}.
Accounting only for first generation quarks and QCD effects, the theoretical moments can be written as~\cite{Braaten:1991qm,LeDiberder:1992jjr,Boito:2014sta,Pich:2016bdg}
\begin{equation}
\label{eq:momdef}
A_{W}(s_0) \, =\,  \frac{N_c}{2} \,S_{\rm ew}\,|V_{ud}|^2 \Big[\,
\delta^{\rm tree}_{W} + \delta^{(0)}_{W}(s_0)  +
\sum_{d\geq 4}\delta^{(d)}_{W}(s_0) \Big] \,,
\end{equation}
where $N_c=3$, $S_{\rm ew}$ stands for electroweak corrections (which we do not consider further), $V_{ud}$ is a CKM matrix element and $s_0$ is the upper bound of the spectral function integration.
The term $\delta^{\rm tree}_{W}$ is the tree-level contribution and $\delta^{(0)}_{W}(s_0)$ stands for the higher order perturbative QCD corrections. The terms $\delta^{(d)}_{W}(s_0)$ represent condensate corrections in the framework of the operator product expansion (OPE)~\cite{Shifman:1978bx}. They involve vacuum matrix elements of low-energy QCD operators of increasing dimension resulting from an expansion in inverse powers of $s_0$. The leading dimension $d=4$ term is related to the well-known gluon condensate $\langle G^{\mu\nu}G_{\mu\nu}\rangle$. There are also so-called duality-violation corrections which are not relevant for the subsequent discussion and therefore suppressed in Eq.(\ref{eq:momdef}). 
Using the 5-loop results~\cite{Gorishnii:1990vf,Surguladze:1990tg,Baikov:2008jh} an impressive precision of about $5\%$ has been achieved for $\alpha_s(m_\tau^2)$ (corresponding to an uncertainty of $1.5\%$ for $\alpha_s(m_Z^2)$), where the uncertainty is dominated by the perturbative error in $\delta^{(0)}_{W}(s_0)$~\cite{Boito:2014sta,Pich:2016bdg,Boito:2020xli,Zyla:2020zbs}. 

A major limitation of $\alpha_s$ determinations from the moments $A_{W}(s_0)$ arises from the fact that two different expansion approaches to evaluate $\delta^{(0)}_{W}(s_0)$, called
contour-improved perturbation theory (CIPT) and fixed-order perturbation theory (FOPT) yield systematic numerical differences that do not seem to be covered by the conventional perturbative uncertainty estimates related to renormalization scale variations. Since CIPT in general leads to smaller values for $\delta^{(0)}_{W}(s_0)$ than FOPT, extractions of $\alpha_s(m_\tau^2)$ based on CIPT generally arrive at larger values than those based on FOPT. 

At this time there is no common agreement in the literature whether this issue is (A) related to the established property that the QCD perturbation series for $\delta^{(0)}_{W}(s_0)$ is asymptotic (which allows for the possibility that the CIPT-FOPT difference is a systematic effect that may be understood and reconciled already based on the available results) or (B)
simply an artifact of a too low truncation order of the series (which would mean that the CIPT-FOPT difference is a better estimator of the perturbative uncertainty than scale variation and the problem may be resolved by calculations beyond 5 loops)\footnote{In the absence of a resolution, this is the conservative attitude adopted in the most recent Review of Particle Physics~\cite{Zyla:2020zbs}.}.
In this context a number of theoretical studies~\cite{Jamin:2005ip,Beneke:2008ad,Caprini:2009vf,DescotesGenon:2010cr,Caprini:2011ya,Beneke:2012vb} were carried out using the renormalon calculus~\cite{Mueller:1984vh,Beneke:1998ui} based on concrete models for the Borel (transformation) function of the Adler function. Their aim was to explore the resulting higher order behavior of the CIPT and FOPT series beyond the concretely known five-loop level and to learn about the possible systematics. An important outcome of these studies was that case (A) is related to Borel function models which contain a sizeable gluon condensate cut, while case (B) is related to Borel function models where a gluon condensate cut is strongly suppressed~\cite{DescotesGenon:2010cr,Beneke:2012vb}. For models with a sizeable gluon condensate cut it was furthermore found that the FOPT series in general approach the model's Borel sum prior to becoming divergent. CIPT seems to approach a value as well, which can, however, be significantly different from the Borel sum. In these studies 
no quantitative or predictive conceptual understanding concerning this discrepancy was gained.

In the recent work~\cite{Hoang:2020mkw} we have provided a quantitative and conceptual understanding. We have shown that the Borel representations (and thus also the Borel sums) for  $\delta^{(0)}_{W}(s_0)$ in the FOPT and in the CIPT expansions are intrinsically different in the presence of infrared (IR) renormalons. We proved that the Borel representation (and the associated Borel sum) known previously for the spectral function moments belongs to the FOPT approach. The CIPT approach, on the other hand, entails a Borel representation that is analytically inequivalent. The resulting difference between the CIPT and FOPT Borel sum, called the asymptotic separation, provides a quantitative explanation of the observations made in Refs.~\cite{Beneke:2008ad,DescotesGenon:2010cr,Beneke:2012vb}. An intriguing implication of these findings is that the OPE corrections that need to be added to  $\delta^{(0)}_{W}(s_0)$ in the CIPT approach differ from those in the FOPT approach and may not have the standard form commonly assumed in previous phenomenological studies. 

In this article we review the findings and considerations given in Ref.~\cite{Hoang:2020mkw}. For the presentation of analytic results we focus on the large-$\beta_0$ approximation, where all expressions can be written down in closed form. The large-$\beta_0$ results have all qualitative features of the full QCD results which were also provided in~\cite{Hoang:2020mkw}, but they are more transparent and less involved.  
We emphasize that the results we report apply to any Borel function model for the Adler function and therefore also to the exact Borel function, if it ever becomes known. The results do, however, by themselves not contribute to the question whether Borel models related to either the cases (A) or (B) mentioned above are more realistic. 
However, only for Borel functions of kind (A) the asymptotic separation and its implications are numerically sizeable, so that they could 
explain the discrepancy observed for the 5-loop CIPT and FOPT spectral function moments.

\section{Theory Input for the $\tau$ Spectral Function Moments}
\label{sec:theory}

In this section we review the theoretical input for the spectral function moments $A_{W}(s_0)$ given in Eq.(\ref{eq:momdef}) and set up our notation. We provide the expressions for $\delta^{(0)}_{W}(s_0)$ in the CIPT and FOPT expansions and discuss the form of the OPE corrections in the established standard approach. We furthermore review the FOPT Borel representation (which was commonly believed to apply to both FOPT and CIPT prior to our work) and present the novel expression for the CIPT Borel representation.  All results are provided in the large-$\beta_0$ approximation, and, when suitable, comments on form of the full QCD expressions are given. 

The QCD corrections  $\delta^{(0)}_{W}(s_0)$ are obtained from the expression ($x\equiv s/s_0$)
\begin{equation}
\label{eq:deltadef}
\delta^{(0)}_{W}(s_0)\, =\,\frac{1}{2\pi i}\,\,\ointctrclockwise_{|s|=s_0}\!\! \frac{{\rm d}s}{s}\,W({\textstyle \frac{s}{s_0}})\,\hat D(s)
\, =\,
\frac{1}{2\pi i}\,\,\ointctrclockwise_{|x|=1}\!\! \frac{{\rm d}x}{x}\,W(x)\,\hat D(x s_0)\,.
\end{equation}
where $\hat D(s)$ is the partonic (reduced) Adler function. It is related to the partonic vaccum polarization function by the relation\footnote{We consider the common definition of parton level results with loop corrections using dimensional regularization and the ${\overline {\rm MS}}$ renormalization scheme unless stated otherwise. We use the convention that the Adler function $D(s)$ and the vacuum polarization function $\Pi(s)$ are real-valued along the Euclidean negative $s$-axis where $\hat D(s)=\hat D(-|s|)$.}
%\begin{equation}
%\label{eq:redAdler}
$\frac{1}{4\pi^2}(1+\hat D(s)) \, \equiv \, -\,s\,\frac{{\rm d}\hat\Pi(s)}{{\rm d} s}$.
%\end{equation} 
The weight function $W(x)$ is a polynomial in $x$ which vanishes for $x=1$ and (together with the choice of $s_0$) specifies the type of moment considered. The contour integral starts/ends at $s=s_0\pm i 0$ ($x=1\pm i0$) and is analytically related to an associated integration along the real positive $s$-axis over the experimental spectral function data~\cite{Pich:2016bdg}. The circular path is convenient, but not mandatory. It may be deformed as long as it stays in the region where the strong coupling remains perturbative.
For $W_\tau(x)=(1-x)^3(1+x)=1-2x+2x^3-x^4$ and $s_0=m_\tau^2$ the moment $A_{W_\tau}(m_\tau^2)$ agrees with the normalized total hadronic $\tau$ decay rate  $R_\tau=\Gamma(\tau^-\to\mbox{hadrons}\,\nu_\tau(\gamma))/\Gamma(\tau^-\to e^-\bar{\nu}_e\nu_\tau(\gamma))$. 

The CIPT expansion~\cite{LeDiberder:1992jjr} for $\delta^{(0)}_{W}(s_0)$ is based on the Adler function series in powers of the complex-valued $\alpha_s(-s)$:
\begin{eqnarray}
\label{eq:AdlerseriesCIPT} 
\hat D(s) & \, =  \,& 
\, \sum_{n=1}^\infty
\bar c_{n} \,a^n(-x)\,, 
\end{eqnarray}
where we define (using $\beta_0=11-2n_f/3$ for the 1-loop $\beta$-function coefficient)
\begin{eqnarray}
\label{eq:adef}
\textstyle 
a(-x)\, \equiv\, \frac{\beta_0\,\alpha_s(-s)}{4\pi}  \, = \, \frac{\beta_0\,\alpha_s(-x s_0)}{4\pi} \,,
\quad
a_0\, \equiv\, \frac{\beta_0\,\alpha_s(s_0)}{4\pi}  \, = \, a(1)\,,
\end{eqnarray} 
with $s_0=|s|$. In the CIPT expansion, the non-analytic structure of the Adler function is entirely encoded by the strong coupling.
The CIPT expansion for the spectral function moments yields the series
\begin{equation}
\label{eq:deltaCIPT}
\delta^{(0),{\rm CIPT}}_{W}(s_0)\, =\, 
\,\frac{1}{2\pi i} \,  \sum\limits_{n=1}^\infty \bar c_{n} \,
\ointctrclockwise_{|x|=1}\!\! \frac{{\rm d}x}{x}\,W(x)\,a^n(-x)\,.
\end{equation}

The FOPT expansion for $\delta^{(0)}_{W}(s_0)$ is based on the Adler function series in powers of the real-valued $\alpha_s(s_0)$, which in the large-$\beta_0$ approximation reads
\begin{eqnarray}
\label{eq:AdlerseriesFOPT}
\hat D(s) & \, =  \,& \,
\sum\limits_{m=1}^\infty\,
a_0^m \, \sum\limits_{i=0}^{m-1} \frac{(m-1)!}{i!(m-i-1)!}\,\bar c_{m-i}\,(-\ln(-x))^{i} \,.
\end{eqnarray}
Here the non-analytic structure of the Adler function is encoded in explicit powers of logarithms.
The FOPT expansion yields the moment series
\begin{equation}
\label{eq:deltaFOPT}
\delta^{(0),{\rm FOPT}}_{W}(s_0)\, =\, 
\,\frac{1}{2\pi i} \, 
\sum\limits_{m=1}^\infty\,
a_0^m \, \sum\limits_{i=0}^{m-1} \frac{(m-1)!}{i!(m-i-1)!}\,\bar c_{m-i}\,\ointctrclockwise_{|x|=1}\!\! \frac{{\rm d}x}{x}\,W(x)\, (-\ln(-x))^{i} 
\,.
\end{equation}

In the (RGI-invariant) standard OPE approach the power correction terms $\delta^{(d)}_{W}(s_0)$ are proportional to scale-independent vacuum matrix elements and perturbative Wilson coefficients times inverse powers of $s_0$. In full QCD their structure is quite complicated due to the large number of operators for $d>4$ and their mixing. In the large-$\beta_0$ approximation, however, all Wilson coefficients are unity, and only two independent operator matrix elements arise for even values of $d>4$ (i.e.\ $d=6,8,\ldots$). Here, the OPE series at the level of the Adler function can be written down in closed form:
\begin{eqnarray}
\label{eq:AdlerOPE}
\hat D^{\rm OPE}(s) & = & 
\frac{1}{(-s)^2} \langle \bar{\cal O}_{4,0}\rangle  + \sum\limits_{p=3}^\infty \frac{1}{(-s)^{p}} \Big[  \langle \bar{\cal O}_{2p,0}\rangle + 
(a(-x))^{-1}  \langle \bar{\cal O}_{2p,-1}\rangle  \Big]\,.
\end{eqnarray}
We have kept the operators generic since their exact form does not matter for the purpose of our discussion.
The anomalous dimension of the operators ${\cal O}_{2p,0}$ (including the gluon condensate) vanish, while the
operators ${\cal O}_{2p,-1}$ have an anomalous dimension proportional to $\beta_0$ such that the leading log resummation yields the prefactor $(a(-x))^{-1}$.
The OPE corrections can thus be written down explicitly: 
\begin{eqnarray}
\label{eq:deltaOPEdef}
\delta^{(4)}_{W}(s_0) & = & \frac{1}{2\pi i}\,\,\ointctrclockwise_{|s|=s_0}\!\! \frac{{\rm d}s}{s}\,W({\textstyle \frac{s}{s_0}})\,\frac{1}{(-s)^2} \langle \bar{\cal O}_{4,0}\rangle\,,  \\ 
\label{eq:deltaOPEdef2}
\delta^{(d=6,8,\ldots)}_{W}(s_0) & = & \frac{1}{2\pi i}\,\,\ointctrclockwise_{|s|=s_0}\!\! \frac{{\rm d}s}{s}\,W({\textstyle \frac{s}{s_0}})\,\frac{1}{(-s)^{d/2}} \Big[  \langle\bar{\cal O}_{d,0}\rangle + 
(a(-x))^{-1}  \langle\bar{\cal O}_{d,-1}\rangle  \Big]\,.
\end{eqnarray}
In concrete phenomenological evaluations the generic scaling $\langle \bar{\cal O}_{d,0/-1}\rangle \sim \Lambda_{\rm QCD}^d$ is assumed, where $\Lambda_{\rm QCD}$ is the QCD hadronization scale\footnote{In the large-$\beta_0$ approximation we define $\Lambda_{\rm QCD}$ as the scale where the solution for the running strong coupling diverges, i.e.\ $a_0=1/\ln(s_0/\Lambda_{\rm QCD}^2)$.}, and the OPE series is frequently terminated at $d=8$~\cite{Pich:1999hc}, see also Ref.~\cite{Boito:2016oam}.
It is an important feature that the terms proportional to the condensates $\langle\bar{\cal O}_{d,0}\rangle$ vanish identically if the polynomial $W(x)$ does not contain a term $x^{d/2}$, as can be easily seen from Eqs.(\ref{eq:deltaOPEdef}) and (\ref{eq:deltaOPEdef2}) using the residue theorem. So the dominant gluon condensate correction and its sizeable parametric uncertainty vanish for $R_\tau$. In full QCD the gluon condensate does not vanish entirely due to the small logarithmic momentum dependence in its Wilson coefficient, but it is still strongly suppressed. In fact, all recent phenomenological $\alpha_s$ determinations were commonly carried out for weight functions that do not contain a term $x^2$ in order to suppress the effects of the gluon condensate and because the CIPT and FOPT perturbative series turn out to be more stable. 

It can be seen from Eqs.(\ref{eq:deltaOPEdef}) and (\ref{eq:deltaOPEdef2}), that
using either the FOPT or the CIPT approach to evaluate $\delta^{(d)}_{W}(s_0)$ in the standard OPE approach only affects the treatment of the contributions of the condensates $\langle{\cal O}_{d,-1}\rangle$. This is due to the momentum dependence of the log-resummed Wilson coefficients. However, these contribute only for $d\ge 6$, so that the resulting difference is extremely small.
In fact, it is smaller than the individual scale variations of $\delta^{(0),{\rm CIPT}}_{W}$ and $\delta^{(0),{\rm FOPT}}_{W}$. So, in practice, the OPE corrections that are added to $\delta^{(0),{\rm CIPT}}_{W}$ and $\delta^{(0),{\rm FOPT}}_{W}$ in the standard approach are identical, and
the discrepancy between the CIPT and FOPT expansions directly affects the $\alpha_s$ determination obtained from both approaches.

That the suppression of OPE corrections is associated to a more stable behavior of the moment's perturbation series is not accidental, but related to the fact that QCD perturbation theory yields series that are asymptotic. In the context of the renormalon calculus~\cite{Gross:1974jv,tHooft:1977xjm,David:1983gz,Mueller:1984vh,Beneke:1998ui} each term in the OPE can be associated to a certain type of order-dependent factorial divergence pattern contained in the perturbative coefficients.  
In the large-$\beta_0$ approximation, where the coefficients $\bar c_n$ have been computed to all orders~\cite{Broadhurst:1992si}
this association can be made very explicit. Within the renormalon calculus this yields an explicit result for the Borel representation of the partonic Adler function for an {\it Euclidean momentum transfer} $s=-s_0$, $\hat D(-s_0)$, with respect to the expansion in $\alpha_s(s_0)$, 
\begin{equation}
\label{eq:invBorelD}
\hat D(-s_0) = \int_0^\infty \!\! {\rm d} u \,\, 
B[\hat D](u)\,e^{-\frac{u}{a_0}}\,,
\end{equation}
where the Borel function has the form~\cite{Broadhurst:1992si}
\begin{eqnarray}
\label{eq:AdlerBorelb0}
\textstyle 
B[\hat D](u)  & =  & \textstyle 
\frac{128}{3\beta_0}\,e^{5u/3}\Big\{ 
\frac{3}{16(2-u)} +\sum\limits_{p=3}^\infty \Big[ \frac{d_2(p)}{(p-u)^2} - \frac{d_1(p)}{p-u} \Big]
-\sum\limits_{p=-1}^{-\infty} \Big[ \frac{d_2(p)}{(u-p)^2} + \frac{d_1(p)}{u-p} \Big]
\Big\}\,,\quad
\end{eqnarray}
with 
$d_2(p) = \frac{(-1)^p}{4(p-1)(p-2)}$ and %\,,\qquad 
$d_1(p) = \frac{(-1)^p(3-2p)}{4(p-1)^2(p-2)^2}$.
The coefficients $\bar c_n$ can be obtained order-by-order from the Taylor series of $B[\hat D](u)$
\begin{equation}
\label{eq:BTaylor}
B[\hat D](u) \, =\, 
\sum\limits_{n=1}^\infty\,  
\frac{u^{n-1}}{\Gamma(n)}\bar c_n \,,
\end{equation}
either by multiplication of the proper $\Gamma(n)$ factors or by using the inverse Borel integral in Eq.(\ref{eq:invBorelD}).
Likewise the form of Eq.(\ref{eq:AdlerBorelb0}) in the entire complex $u$-plane can be uniquely determined from summing the {\it absolute convergent} Taylor series~(\ref{eq:BTaylor}) within its radius of convergence (which is $1$ due to the $p=-1$ poles) and using analytic continuation. The poles for negative $p$ are the so-called UV renormalons and are related to sign-alternating divergence patterns which can be formally summed in the Borel representation~(\ref{eq:invBorelD}). In QCD the UV renormalons arise from the large-momentum behavior of the loop diagrams. 

The poles for positive $p$ are the so-called the IR renormalons and their  
ambiguities in Eq.(\ref{eq:invBorelD}) quantify the equal-sign divergence patterns in the $\bar c_n$ for large values of $n$. In QCD the IR renormalons arise from the small-momentum behavior of the loop diagrams.
For simplicity, we call the function $B[\hat D](u)$ the 'Adler function's Borel function'. But, in order to avoid confusion, we reiterate that it is the Borel function for the Euclidean and real-valued Adler function $\hat D(-s_0)$ with respect to 
the expansion in the real-valued $\alpha_s(s_0)$. This corresponds to the point $x=-1$, where the 
the two expansions of Eqs.(\ref{eq:AdlerseriesCIPT}) and (\ref{eq:AdlerseriesFOPT}) agree. 
The $s_0$-dependence of the ambiguities arising in the Borel integration of Eq.(\ref{eq:invBorelD}) precisely matches the $s_0$-dependence
of the OPE corrections in $\hat D^{\rm OPE}(-s_0)$. Single poles at $u=p$ are related to the condensates 
$\langle\bar{\cal O}_{2p,0}\rangle$ and double poles to the condensates $\langle\bar{\cal O}_{2p,-1}\rangle$:
\begin{equation}
\label{eq:OPEBorel}
\langle\bar{\cal O}_{2p,0}\rangle \, \Leftrightarrow \, \frac{1}{(p-u)} \qquad {\rm and} \qquad
\langle\bar{\cal O}_{2p,-1}\rangle \, \Leftrightarrow \, \frac{1}{(p-u)^2} \,.
\end{equation}
Within perturbation theory the value of the condensate compensate, order by order, the factorial divergence patterns due to their associated IR renormalon. So within perturbation theory, the condensates are order-dependent quantities similar to the pole mass of a heavy quark~\cite{Hoang:2020iah}.

Upon contour integration over $s$ with a proper weight function, the expression in Eq.(\ref{eq:invBorelD}) with the replacement  $a_0\to a(-x)$ (i.e.\ $\alpha_s(s_0)\to\alpha(-s)$)  immediately leads to the
spectral function moment Borel representation 
\begin{equation}
\label{eq:BorelFOPT}
\delta_{W,{\rm Borel}}^{(0),{\rm FOPT}}(s_0) = {\rm PV} \int_0^\infty \!\! {\rm d} u \,\, 
\frac{1}{2\pi i}\,\ointctrclockwise_{|x|=1} \frac{{\rm d}x}{x} \, W(x) \,
B[\hat D](u)\,e^{-\frac{u}{a(-x)}}\,,
\end{equation}
where PV stands for the commonly adopted principal value prescription (defined by taking the average of deforming the $u$-integration path below and above the real $u$-axis) to obtain a well-defined expression for the Borel sum. Instead of using the order-dependent approach to the OPE just mentioned above, one can in principle also use the Borel sum as the prediction of perturbation theory. In this Borel-resummed approach, the OPE corrections are constants whose values depend on the regularization prescription to render the integration of the Borel representation well-defined. This approach can, however, only be carried out in a model-independent way, if the Borel function is known.

Prior to the results of our work Eq.(\ref{eq:BorelFOPT}) has been the commonly accepted Borel representation for the 
perturbative spectral function moments for the CIPT as well as the FOPT approach. It was also used as the definition for the Borel sum
in the previously mentioned studies~\cite{Jamin:2005ip,Beneke:2008ad,Caprini:2009vf,DescotesGenon:2010cr,Caprini:2011ya,Beneke:2012vb}.
The simple argument to support this view was that inserting 
the Taylor expansion~(\ref{eq:BTaylor}) and pulling out the sum over $n$ provides the form of the $\delta^{(0),{\rm CIPT}}_{W}(s_0)$ series
in Eq.(\ref{eq:deltaCIPT}). On the other hand, expanding that series further in powers of $\alpha_s(s_0)$ obviously leads to the FOPT series $\delta^{(0),{\rm FOPT}}_{W}(s_0)$ in Eq.(\ref{eq:deltaFOPT}). 
However, this argumentation only says that CIPT and FOPT are two different expansion approaches to the same underlying (asymptotic) series. This is undoubtedly true. But this is not the point when it comes to properties of the Borel representation and the Borel sum, where renormalon structures beyond the plain perturbation series may become relevant. 
In Ref.~\cite{Hoang:2020mkw} it was shown that Eq.(\ref{eq:BorelFOPT}) is the Borel representation for the FOPT spectral function moments, but it does not apply for CIPT. We therefore already added the superscript FOPT in $\delta_{W,{\rm Borel}}^{(0),{\rm FOPT}}(s_0)$.  

For CIPT spectral function moments the correct Borel representation has the form
\begin{equation}
\label{eq:BorelCIPT}
\delta_{W,{\rm Borel}}^{(0),{\rm CIPT}}(s_0) = \int_0^\infty \!\! {\rm d} \bar u \,\,  
\frac{1}{2\pi i}\,\ointctrclockwise_{{\cal C}_x} \frac{{\rm d}x}{x} \, W(x) \,
\big({\textstyle \frac{a(-x)}{a_0}}\big)\,
B[\hat D]\Big({\textstyle \frac{a(-x)}{a_0}} \bar u\Big)
\,e^{-\frac{\bar u}{a_0}}\,.
\end{equation}	
At the Euclidean point $x=-1$ the integrands in Eqs.(\ref{eq:BorelFOPT}) and (\ref{eq:BorelCIPT}) agree, but the Borel representations differ due to the integration in the complex $x$ plane. Since in $\delta_{W,{\rm Borel}}^{(0),{\rm CIPT}}(s_0)$ the complex-valued coupling appears inside the argument of the Borel function $B[\hat D]$, the contour $x$-integral becomes aware of the non-analytic renormalon structures contained in $B[\hat D](u)$. Thus special care is needed for the path ${\cal C}_x$ such that it does not cross cuts coming from IR renormalons.
The path ${\cal C}_x$ still starts/ends at $x_{\mp}=1\pm i0$ as for $\delta_{W,{\rm Borel}}^{(0),{\rm FOPT}}(s_0)$, but in general needs to be deformed further into the negative real complex plane, as we discuss in more detail in Secs.~\ref{sec:analytic} and \ref{sec:asymptoticseparation}.

Both Borel representations are related via the {\it complex-valued} change of variables $u=\bar u \,\alpha_s(-s)/\alpha_s(s_0)=\bar u\, a(-x)/a_0$. As a consequence, a difference between $\delta_{W,{\rm Borel}}^{(0),{\rm FOPT}}(s_0)$ and $\delta_{W,{\rm Borel}}^{(0),{\rm CIPT}}(s_0)$ arises due to the presence of the non-analytic IR renormalon singularities in the Borel function $B[\hat D](u)$. Moreover, due to the finite imaginary part of $\alpha_s(-s)$ along the contour integration the $\bar u$-integration path never touches the non-analytic IR renormalon structures in  $B[\hat D](u)$ located  along the positive $u$-axis. The $\bar u$-integration is therefore automatically regularized, in contrast to $\delta_{W,{\rm Borel}}^{(0),{\rm FOPT}}(s_0)$, where a regularization such as the PV prescription must always be imposed by hand. We come back to this issue in Sec.~\ref{sec:Adlerfunction}. As it turns out, the difference $\delta_{W,{\rm Borel}}^{(0),{\rm CIPT}}(s_0)-\delta_{W,{\rm Borel}}^{(0),{\rm FOPT}}(s_0)$ -- which we call the {\it asymptotic separation} -- can by far exceed the ambiguity of the FOPT Borel sum (commonly defined from the difference of deforming the $u$-integration path in Eq.(\ref{eq:BorelFOPT}) below and above the positve real $u$-axis), if the Borel function contains a sizeable gluon condensate ($p=2$) IR singularity. In the large-$\beta_0$ approximation the residue of the gluon condensate pole is quite sizeable as can be seen from Eq.(\ref{eq:AdlerBorelb0}).

\section{Proof for the FOPT and CIPT Borel Representations}
\label{sec:representations}

Let us now review the proof that Eqs.(\ref{eq:BorelFOPT}) and (\ref{eq:BorelCIPT}) are the correct Borel representations for the FOPT and CIPT spectral function moment series, respectively. 

We first consider CIPT and start from the observation that the results for the contour integrals over the product of functions $\frac{1}{x} W(x)(\frac{\alpha_s(-x s_0)}{\pi})^n$ in Eq.(\ref{eq:deltaCIPT}) involve a nontrivial interplay of the $x$-dependences of the weight function $W(x)$ and the functions $\alpha_s^n(-x s_0)$. This shows that the resulting CIPT spectral function moment series may not be considered as an expansion in $\alpha_s(-x s_0)$ and that the concrete numbers obtained from the contour integrals should be considered as part of the series coefficients. A more appropriate expansion parameter is obviously $\alpha_s(s_0)$, which one can conveniently pull out of the series coefficients with the appropriate power,
\begin{equation}
\label{eq:deltaCIPT2}
\delta^{(0),{\rm CIPT}}_{W}(s_0)\, =\, 
\,\frac{1}{2\pi i} \,  \sum_{n=1}^\infty \bar c_{n} \,\Big[\,
\ointctrclockwise_{|x|=1}\!\! \frac{{\rm d}x}{x}\,W(x)\,\Big({\textstyle \frac{a(-x)}{a_0}}\Big)^n\,\Big]\,
a_0^n\,,
\end{equation}
(In fact, one can pull out powers of any definite small real constant, which is equivalent to this choice via a {\it real-valued} rescaling of $u$ that leaves the Borel integral unchanged.) 
So the CIPT series is generated by the perturbative Borel representation 
\begin{equation}
\label{eq:BorelCIPTTaylor}
\delta_{W,{\rm Borel}}^{(0),{\rm CIPT}}(s_0) = \int_0^\infty \!\! {\rm d} \bar u \,\,  
 \sum_{n=1}^\infty  \,\Big[\,  \frac{1}{2\pi i}\,\ointctrclockwise_{|x|=1} \frac{{\rm d}x}{x} \, W(x) \,
\frac{\bar c_{n}}{\Gamma(n)}\,\Big({\textstyle \frac{a(-x)}{a_0}}\Big)^n\,
\bar u^{n-1} 
\,\Big]
\,e^{-\frac{\bar u}{a_0}}\,.
\end{equation}
We can now obtain the full CIPT Borel function by summing the $\bar u$-Taylor series and using analytic continuation. Recalling the convergence properties of the $B[\hat D](u)$ Taylor series~(\ref{eq:BTaylor}) in the complex $u$-plane we see that 
the $\bar u$-series in Eq.(\ref{eq:BorelCIPTTaylor}) is absolute convergent for $|\bar u|<1$ as well\footnote{As explained in Sec.~\ref{sec:analytic} the convergence radius is actually $|1+i a_0\pi|$.}, since $|a(-x)/a_0|=|\alpha_s(-x s_0)/\alpha_s(s_0)|\le 1$ along the countour with $|x|=1$ (and in fact along any contour with $|x|\ge 1$). The $\bar u$-series thus uniquely defines  
a Borel function of the form $(\frac{a(-x)}{a_0})B[\hat D]((\frac{a(-x)}{a_0})\bar u)$ in the entire complex $\bar u$ plane and for any complex $x$ along the contour. This Borel function inherits all non-analytic structures already contained in $B[\hat D](u)$. We thus obtain Eq.(\ref{eq:BorelCIPT}) as the correct Borel representation for the CIPT spectral function moments. The proof does actually not at all rely on the large-$\beta_0$ approximation and applies in principle in the same manner in full QCD (where, of course, the exact expression for $B[\hat D](u)$ is unknown).  

Let us now come to the proof that Eq.(\ref{eq:BorelFOPT}) is the correct Borel representation for the FOPT moments. At the first glance this does not appear obvious: Since $\alpha_s(-s)$ appears explicitly in the exponential, the CIPT interpretation appears plausible. However, Eq.(\ref{eq:BorelFOPT}) can be rewritten {\it without changing the Borel integral} such, that its association with the FOPT series is made imperative. To this end we reexpand $\alpha_s(-s)$ in the exponential in terms of $\alpha_s(s_0)$, which by itself is a {\it convergent} manipulation for sufficiently large $s_0$. The expansion does therefore not change the integral.
We can now write down an equivalent Borel representation with respect to $\alpha_s(s_0)$:
\begin{equation}
\label{eq:BorelFOPTrewr}
\delta_{W,{\rm Borel}}^{(0),{\rm FOPT}}(s_0) = {\rm PV}\int_0^\infty \!\! {\rm d} u \,\, 
\frac{1}{2\pi i}\,\ointctrclockwise_{|x|=1} \frac{{\rm d}x}{x} \, W(x) \,
B[\hat D^{\rm FOPT}](u,\ln(-x))\,e^{-\frac{u}{a_0}}\,.
\end{equation}
In the large-$\beta_0$ approximation, where we have $1/a(-x) = 1/a_0 +\ln(-x)$ this expansion is actually quite trivial. The resulting form of $B[\hat D^{\rm FOPT}](u,\ln(-x))$ is 
\begin{eqnarray}
\label{eq:BorelfctFOPT}
B[\hat D^{\rm FOPT}](u,\ln(-x)) & \equiv & B[\hat D](u)\,e^{-\frac{u}{a(-x)}+\frac{u}{a_0}} = B[\hat D](u) \,  e^{-u\ln(-x)}\,.
\end{eqnarray}
The Taylor expansion in powers of $u$ can be obtained in a straightforward way and reads $B[\hat D^{\rm FOPT}](u,\ln(-x))=\sum_{m=1}^\infty u^{m-1}
\sum_{i=0}^{m-1} \frac{\bar c_{m-i} (-\ln(-x))^{i}}{i!(m-i-1)!}$. It is thus immediately 
obvious that Eq.(\ref{eq:BorelFOPTrewr}) is indeed the Borel representation of the FOPT series in Eq.(\ref{eq:deltaFOPT}). The essential point of our manipulation was that for the derivation of $B[\hat D^{\rm FOPT}](u,\ln(-x))$ in Eq.(\ref{eq:BorelfctFOPT}) we {\it did not rely on the $u$-Taylor expansion}, which ensures that the Borel representation (i.e.\ its Borel sum) is not modified. In the context of full QCD, the derivation of the form of $B[\hat D^{\rm FOPT}](u,\ln(-x))$ is actually a bit more involved since the expansion of the $1/\alpha_s(-s)$ term in the exponent does not terminate. But the expansion is convergent, and one can use partial integration (which does not rely on the Taylor series for $B[\hat D](u)$ and does not change the Borel sum) to derive $B[\hat D^{\rm FOPT}](u,\ln(-x))$. We refer to  Ref.~\cite{Hoang:2020mkw} for details.

Let us summarize: The different form of the Borel representations $\delta_{W,{\rm Borel}}^{(0),{\rm CIPT}}(s_0)$ and $\delta_{W,{\rm Borel}}^{(0),{\rm FOPT}}(s_0)$ arises directly from the different character of the CIPT and FOPT moment series when computing their respective Borel functions. So the difference arises from an {\it explicit calculation}. The different (non-)analytic properties of $\delta_{W,{\rm Borel}}^{(0),{\rm CIPT}}(s_0)$ and $\delta_{W,{\rm Borel}}^{(0),{\rm FOPT}}(s_0)$, that eventually lead to the asymptotic separation, become manifest after the ($u$ or $\bar u$) Taylor series are resummed and analytically continued to obtain the Borel functions in the entire complex ($u$ or $\bar u$) Borel planes. These procedures are unique and unambiguous. It is the latter aspect that is the essential point. If we would rely only on the ($u$ or $\bar u$) Taylor series, we would immediately arrive at the equivalence of  $\delta_{W,{\rm Borel}}^{(0),{\rm CIPT}}(s_0)$ and $\delta_{W,{\rm Borel}}^{(0),{\rm FOPT}}(s_0)$, but then we were back to the old naive argumentation mentioned at the end of the previous section.

\section{Analytic Examination of the FOPT and CIPT Borel Functions}
\label{sec:analytic}

Let us now have a closer look at the analytic properties of the FOPT and CIPT moments' Borel functions that arise after carrying out the contour integrations.  
The properties of the resulting Borel functions for the FOPT moments are well known, but we discuss them here as well to better illustrate the differences to the CIPT case.
Since the differences only arise from IR renormalon singularities in the Borel function $B[\hat D](u)$ and in order to keep the examination simple, we consider 
a single generic IR renormalon term contained in Eq.(\ref{eq:AdlerBorelb0}) of the form
 \begin{equation}
\label{eq:BorelDir}
B^{\rm IR}_{\hat D,p,\gamma}(u) \, = \, \frac{1}{(p-u)^\gamma}
\end{equation}
and a  monomial weight function $W(x)=(-x)^m$. 

We start with the FOPT case.
In the large-$\beta_0$ approximation it is straightforward to carry out the contour integration~\cite{Ball:1995ni}:
\begin{eqnarray}
\nonumber
\delta_{\{(-x)^m,p,\gamma\},{\rm Borel}}^{(0),{\rm FOPT}}(s_0) & = & {\rm PV}  \int_0^\infty \!\! {\rm d}u \, 
\frac{e^{-\frac{u}{a_0}}}{(p-u)^\gamma}\,
\frac{1}{2\pi i}\,\ointctrclockwise_{|x|=1} \frac{{\rm d}x}{x} \,(-x)^m \,
e^{-u\ln(-x)}\\
\label{eq:BorelFOPT2} 
& = & {\rm PV}  \int_0^\infty \!\! {\rm d}u \, 
\,\frac{(-1)^m \sin(u\pi)}{\pi(u-m)}\,  \frac{e^{-\frac{u}{a_0}}}{(p-u)^\gamma}	\,.
\end{eqnarray} 
The resulting modulation factor $\sin(u\pi)/(u-m)$ has zeros at integer values for $u$, except for $u=m$.
For $p\neq m$ it completely eliminates the single poles (and their associated renormalon ambiguity) and reduces the double poles to single poles.
Thus, we can drop the PV prescription for single poles in generic moments with $p\neq m$, and the associated
contributions in the FOPT moment's series become {\it convergent}. This goes hand-in-hand with the 
elimination of the corresponding standard OPE correction associated with the condensate $\langle\bar {\cal O}_{2p,0}\rangle$ already mentioned above Eq.(\ref{eq:OPEBorel}).
In full QCD, due to the higher-order terms in the QCD $\beta$-function and because the non-analytic renormalon structures of the Borel function  $B[\hat D](u)$ are cuts rather than poles, a complete elimination does not take place, but it is well-known that the corresponding renormalons and OPE corrections still become strongly suppressed~\cite{Beneke:1998ui}. 
Note that the analogue of Eqs.(\ref{eq:BorelFOPT2}) in full QCD has been computed analytically in Ref.~\cite{Hoang:2020mkw}.     

The phenomenological importance of this cancellation (or suppression) is that it applies to the gluon condensate renormalon contribution. This is the theoretical explanation for the observation that spectral function moment series are more stable if $W(x)$ does not contain a $x^2$ term. It is also compatible with the vanishing (or suppression) of the gluon condensate OPE correction mentioned after Eq.~(\ref{eq:deltaOPEdef2}). 

Let us now consider the CIPT case. The corresponding generic terms in the Borel representation adopt the form
\begin{eqnarray}
\label{eq:BorelCIPT3}
\delta_{\{(-x)^m,p,\gamma\},{\rm Borel}}^{(0),{\rm CIPT}}(s_0) & = &
\int_0^\infty \!\! {\rm d} \bar u \, 
\frac{1}{2\pi i}\,\ointctrclockwise_{{\cal C}_x} \frac{{\rm d}x}{x} \,(-x)^m \,
\big({\textstyle \frac{a(-x)}{a_0}}\big)\,
\frac{e^{-\frac{\bar u}{a_0}}}{\big(p-\frac{a(-x)}{a_0}\bar u\big)^\gamma}
\\ \nonumber & = &
\int_0^\infty \!\! {\rm d} \bar u \,  e^{-\frac{\bar u}{a_0}}\,
\tilde C(p,\gamma,m,s_0;\bar u)\,.
\end{eqnarray}
%In the second line we have changed the integration variable to $t=-1/(2a(-x))$ with $-x=(\Lambda_{\rm QCD}^2/s_0)e^{-2t}$ and $t_0\equiv-1/(2a_0)$ for better visualization.\footnote{The $t$-variable notation~\cite{Hoang:2008yj} is quite useful for complete analytic calculations in full QCD.} The $t$-integration path starts/ends at $t_\mp = -1/(2a(-1\mp i0))=\frac{1}{2}(-\frac{1}{a_0}\pm i\pi)$.
Recalling that $a(-x)/a_0=1/(1+a_0\ln(-x))$ we can see that, apart from the Landau pole and the cut along the positive real $x$ axis, which arise from the strong coupling function, there is an additional pole on the negative real $x$-axis where $\alpha_s(-x s_0)= p\alpha_s(s_0)/\bar u$. This pole is located at $\tilde x(\bar u)=- e^{(\bar u-p)/p a_0} = -(\Lambda_{\rm QCD}^2/s_0)^{(p-\bar u)/p}$. In full QCD this pole corresponds to a cut that stretches along the real $x$-axis for $x>\tilde x_{\rm QCD}(\bar u)\approx \tilde x(\bar u)$. For $\bar u<p$ this cut is still within the circular path $|x|=1$. However, for $\bar u>p$ it is not, so that we have to deform the integration path ${\cal C}_x$ further into the negative real $x$-plane such that it crosses the real negative axis at some $x < \tilde x(\bar u)$  (or $x<\tilde x_{\rm QCD}(\bar u)$ in full QCD).
This property entails that, when $\bar u$ increases, the allowed region where the path can cross the negative real $x$-axis is shifted towards negative infinity. Furthermore, when the $\bar u$ integration is carried out first, the contour ${\cal C}_x$ must be deformed to minus negative real infinity. This additional cut will be an essential element for the discussions in Sec.~\ref{sec:Adlerfunction}.

In the large-$\beta_0$ approximation (with $\gamma=1,2$) it is possible to carry out the integration in Eq.(\ref{eq:BorelCIPT3}) analytically to obtain the CIPT moments' Borel function:
\begin{eqnarray}
\label{eq:Cfunc1}
\tilde C(p,1,m,s_0;\bar u) & = & \frac{2\,t_0}{|p|}\, Q\Big(1,m,-2t_0(1-{\textstyle\frac{\bar u}{p}})\Big)\,,
\\ 
\label{eq:Cfunc2}
\tilde C(p,2,m,s_0;\bar u) & = &\frac{2\,t_0}{p^2}\,\Big[ 
Q\Big(1,m,-2t_0(1-{\textstyle\frac{\bar u}{p}})\Big)  
-\frac{2 \bar{u} t_0}{p}\, Q\Big(2,m,-2t_0(1-{\textstyle\frac{\bar u}{p}})\Big)  
\Big],\,\,
\end{eqnarray}
where ($n=1,2,3,\ldots$)
\begin{eqnarray}
\label{eq:Qfunc}
Q(1,0,\rho) & = & \frac{i}{2\pi}\,\Big[\, \ln(\rho+i\pi) -  \ln(\rho-i\pi) \, \Big]\,,
\\ 
Q(n \ge 2,0,\rho) & = & -\frac{i}{2\pi(n-1)}\,\Big[\, (\rho+i\pi)^{1-n} - (\rho-i\pi)^{1-n} \, \Big]\,,
\\ \nonumber
Q(n,m,\rho) & = & m^{n-1}\,e^{-m\rho}\,\left[\,\frac{(-1)^n\,i}{2\pi}\Big(
\Gamma(1-n,-m(\rho+i\pi)) \right.\\
& &\qquad \qquad \left.-\Gamma(1-n,-m(\rho-i\pi))
\Big) -\frac{1}{\Gamma(n)} \, \right]\,. 
\end{eqnarray}
It is straightforward to check that the $\bar u$-Taylor expansion of the $\tilde C$ functions agrees with the terms in Eq.(\ref{eq:BorelCIPTTaylor}) after carrying out the $x$-contour integration. A closer inspection of the $\tilde C$ functions also reveals that for all $p$ and $m$ values they contain poles and cuts (located parallel to the real $\bar u$-axis)
with distance $p|1+i a_0\pi|$ from the origin at $\bar u=0$. These non-analytic structures originate from the poles of the generic Borel function $B^{\rm IR}_{\hat D,p,\gamma}(u)$ in Eq.(\ref{eq:BorelDir}). Thus the $\bar u$-Taylor series converges absolutely for $\bar u<p|1+i a_0\pi|$.

The location of the cuts and poles in the $\tilde C$ functions away from the real $\bar u$-axis entails that the $\bar u$-integral is well-defined without any additional regularization prescription. This property is inherited from the original form of the CIPT Borel representation in Eq.(\ref{eq:BorelCIPT}) that was already pointed out at the end of Sec.~\ref{sec:theory}. However, there is more to it, since in contrast to the FOPT case, the contour integration for CIPT is never ever capable of eliminating the single IR renormalon pole in Eq.~(\ref{eq:BorelDir}). This and the finite radius of convergence of the $\bar u$-Taylor series imply (in the large-$\beta_0$ approximation) that the CIPT moment series are {\it divergent even for a single IR renormalon pole and for $p\neq m$}.\footnote{The existence of cuts or poles in the Borel function implies the divergence of the underlying perturbation series regardless of whether they are located on the positive real axis or not.} We demonstrate this explicitly in our numerical analysis in Sec.~\ref{sec:asymptoticseparation}.
Since for a single IR renormalon pole and $p\neq m$ the associated FOPT moment series is convergent and the OPE corrections involving the condensate $\langle{\cal O}_{2p,0}\rangle $ vanish, one must draw an intriguing conclusion: In the large-$\beta_0$ approximation the OPE corrections for the CIPT moments do in general no have the standard form described in 
Eqs.(\ref{eq:deltaOPEdef}) and (\ref{eq:deltaOPEdef2}).  

In the next section we arrive at the same conclusion using a different and more general line of reasoning concerning the Adler function.
We refer to Ref.~\cite{Hoang:2020mkw} for the discussion of a number of other features of the functions $\tilde C$ functions and of an additional technical subtlety for the case $m\ge p$.

\section{FOPT and CIPT for the Adler Function and the OPE}
\label{sec:Adlerfunction}

In this section we discuss the implications of the FOPT and CIPT expansions for the Adler function. 
The results are the basis for the computation of the asymptotic separation (which can actually also be determined from the results given in the previous section) and they shed further light on the OPE corrections for the CIPT moments and the corresponding expansion of the Adler function in powers of $\alpha_s(-s)$.

\begin{figure} 
	\centering
	\subfloat[\label{fig:u-contourUV}]{\includegraphics[width= 0.49\textwidth]{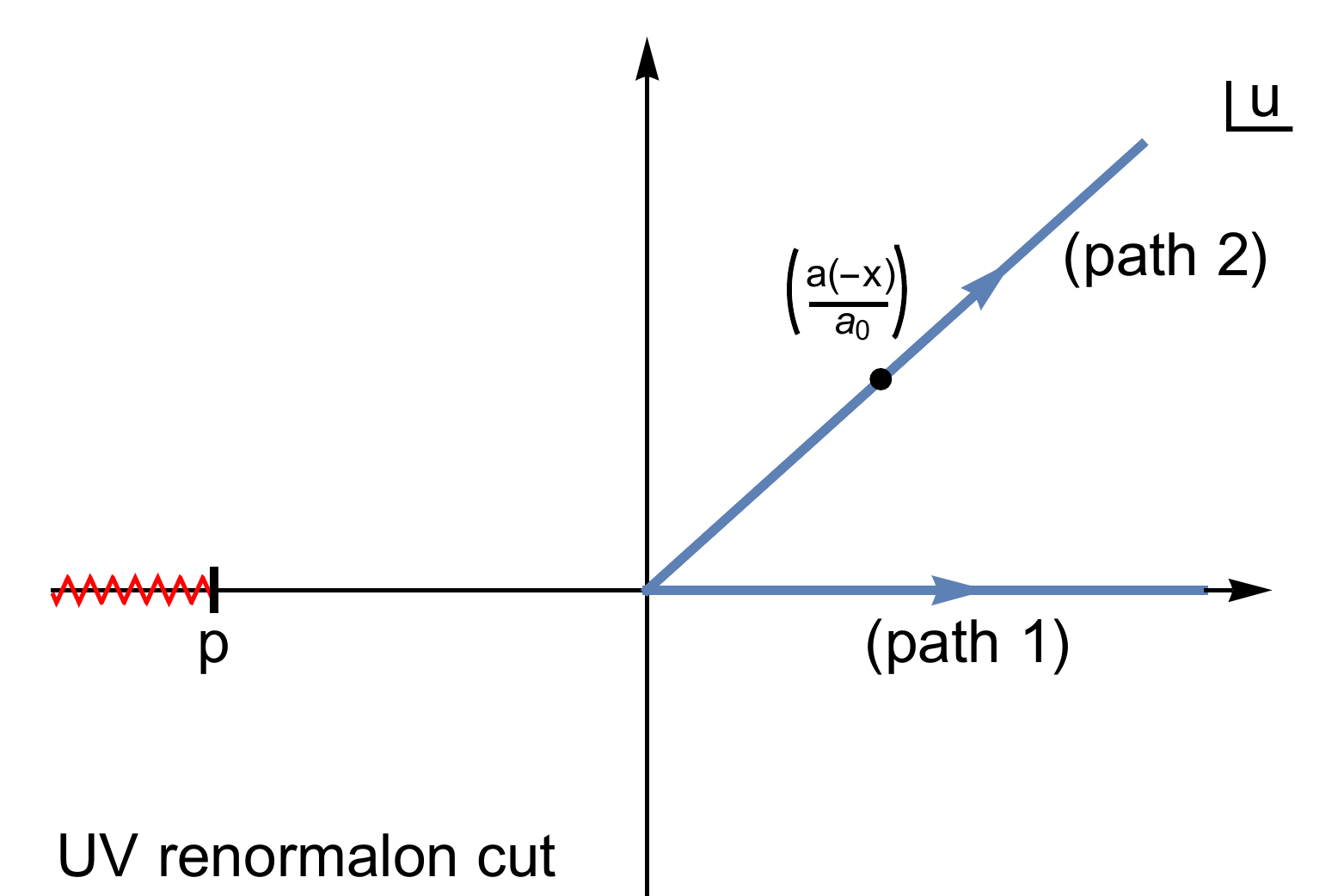}} ~~
	\subfloat[\label{fig:u-contourIR}]{\includegraphics[width= 0.49\textwidth]{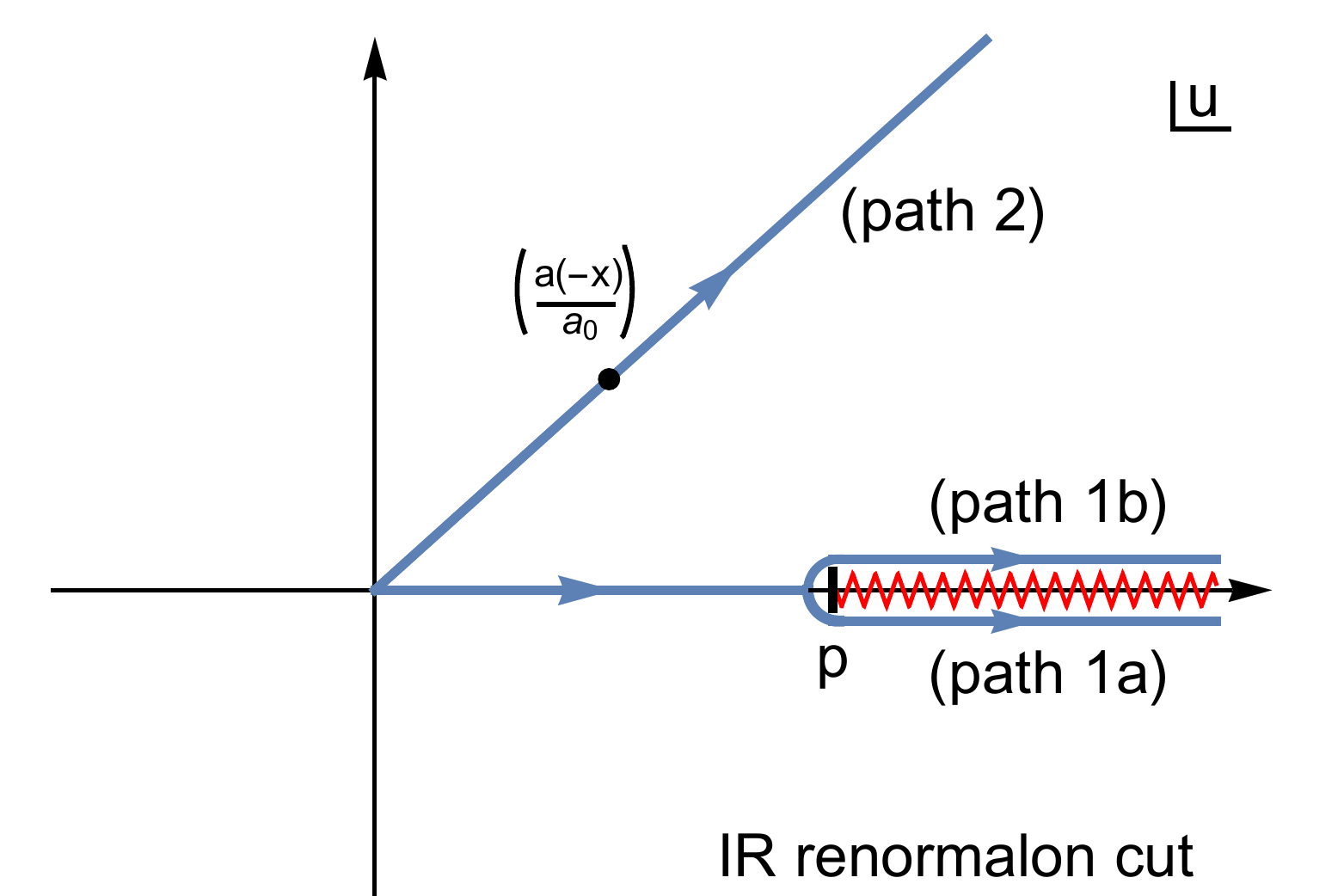}}
	\caption{\label{fig:ucontour} 
		Graphical illustration of the Borel integration paths involved for the FOPT and CIPT Borel representations for the cases of a UV renormalon with $p<0$ (left panel) and an IR renormalon with $p>0$ (right panel). The red zig-zag line represents $u=p$ renormalon cuts in full QCD, which reduce to poles at $u=p$ in the large-$\beta_0$ approximation.}
\end{figure}

Since in the regularized FOPT and CIPT moment Borel representations 
the Borel integrals can be carried out prior to the contour integration, the following
two Borel representations of the Adler function are implied: 
\begin{eqnarray}
\label{eq:AdlerBorelCIPT}
\hat D_{\rm Borel}^{\rm CIPT}(s) & = &\,\int_0^\infty \!\! {\rm d} \bar u \,\,
\big({\textstyle \frac{a(-x)}{a_0}}\big)\,
B[\hat D]\big({\textstyle \frac{a(-x)}{a_0}} \bar u\big)\,e^{-\frac{\bar u}{a_0}}\,, \\
\label{eq:AdlerBorelFOPT}
\hat D_{\rm Borel}^{\rm FOPT}(s) & = &\,{\rm PV}\int_{0}^\infty \!\! {\rm d} u \,\,  B[\hat D](u)\,e^{-\frac{u}{a(-x)}} \,.
\end{eqnarray}
We call them the CIPT and FOPT Borel representations of the Adler function, respectively, just due to their connection to the CIPT and FOPT moment series. We remind the reader that the two Borel representations are relevant for complex $s$ values away from the Euclidean axis, i.e. for $s\neq -s_0=-|s|$ or $x\neq -1$. Along the Euclidean axis where $x=-1$ only the FOPT Borel representation applies. Let us now analyze both Borel representations. 

For analytic terms in $B[\hat D](u)$ as well as for UV renormalon terms of the form $1/(\tilde p + u)^\gamma$ (with $\tilde p>0$) the integrals over $u$ or $\bar u$ both yield identical results. This is because 
the {\it complex-valued} change of variables $u=\bar u \,\alpha_s(-s)/\alpha_s(s_0)=\bar u\, a(-x)/a_0$ that formally relates Eqs.(\ref{eq:AdlerBorelFOPT}) and (\ref{eq:AdlerBorelCIPT})
can be associated to carrying out the $u$ integral either along path~1 or along path~2 shown in Fig.~\ref{fig:u-contourUV} (for the case that the imaginary part of $x$ is positive). Due to the absence of non-analytic structures in the positive complex $u$ plane (recall that the real part of $\alpha_s(-s)$ is positive) both paths lead to identical results, because closing the paths~1 and 2 at positive real infinity does not encircle any non-analytic structures. For a generic UV renormalon pole term the result reads
\begin{eqnarray}
\label{eq:UVBorelInt}
\int_0^\infty \!\! {\rm d} \bar u \,
\frac{({\textstyle\frac{a(-x)}{a_0}})e^{-\frac{\bar u}{a_0}}}{\big(\tilde p + \frac{a(-x)}{a_0} \bar u\big)^\gamma}
= 
\int_0^\infty \!\! {\rm d} u \, 
\frac{e^{-\frac{u}{a(-x)}}}{(\tilde p + u)^\gamma}
= 
(a(-x))^{1-\gamma}\,e^{\frac{\tilde p}{a(-x)}}\,\Gamma\Big(1-\gamma,{\textstyle\frac{\tilde p}{a(-x)}}\Big).
\end{eqnarray}
The expression is analytic in the entire complex $s$-plane except for the Landau pole of the strong coupling and the cut along the positive real $s$-axis (recall that $s=x s_0$). These non-analytic structures arise entirely from the strong coupling function. This also applies when the analytic contributions in $B[\hat D](u)$ are considered, i.e.\ terms that are related to the convergent contributions in the Alder function series. This is the analytic property one expects from the physical Adler function. 

For a generic IR renormalon term in $B[\hat D](u)$ of the form of Eq.(\ref{eq:BorelDir}) the story is quite different. The PV prescription for the FOPT Borel representation 
in Eq.(\ref{eq:AdlerBorelFOPT}) involves the average of the paths~1a and 1b (see Fig.~\ref{fig:u-contourIR}), and the integration for the CIPT Borel representation
is related to path~2 due to the complex-valued change of variables $u=\bar u\, a(-x)/a_0$ (shown for ${\rm Im}[x]>0$ in Fig.~\ref{fig:u-contourIR}). Closing the contours of path~2 with either 1a or 1b at positive infinity thus always encloses the IR poles or cuts located on the positive real $u$-axis, which results in a difference between
$\hat D_{\rm Borel}^{\rm CIPT}(s)$ and $\hat D_{\rm Borel}^{\rm FOPT}(s)$. This difference changes sign depending on the sign of the imaginary part of $x$. 
The results for $\hat D_{\rm Borel}^{\rm CIPT}(s)$ and $\hat D_{\rm Borel}^{\rm FOPT}(s)$ can be readily calculated and read
\begin{eqnarray}
\label{eq:IRBorelDCIPTsum}
\hat D_{p,\gamma,{\rm Borel}}^{\rm CIPT}(s) & =  &
-(-a(-x))^{1-\gamma}\,e^{-\frac{p}{a(-x)}}\,\Gamma\Big(1-\gamma,{\textstyle - \frac{p}{a(-x)}}\Big)\,,\\
\label{eq:IRBorelDFOPTsum}
\hat D_{p,\gamma,{\rm Borel}}^{\rm FOPT}(s) & = & 
 -(-a(-x))^{1-\gamma}\,e^{-\frac{p}{a(-x)}}\,\Gamma\Big(1-\gamma,{\textstyle - \frac{p}{a(-x)}}\Big) \\ \nonumber & &
 - {\rm sig}[{\rm Im}[x]]\,(i\pi)\,\frac{(a(-x))^{1-\gamma}}{\Gamma(\gamma)}\,e^{-\frac{p}{a(-x)}}
\,,
\end{eqnarray}
for any real $\gamma$ so that they are also valid beyond the large-$\beta_0$ approximation. The second term in Eq.(\ref{eq:IRBorelDFOPTsum}) that is proportional to the sign-function ${\rm sig}[{\rm Im}[x]]$ is related to the enclosed cut (or pole). The FOPT expression in Eq.(\ref{eq:IRBorelDFOPTsum}) is analytic in the entire $s$-plane except for the Landau pole of the strong coupling and the cut along the positive real $s$-axis, just like Eq.(\ref{eq:UVBorelInt}). To achieve this analytic behavior the term proportional to ${\rm sig}[{\rm Im}[x]]$ is crucial as it cancels a cut along the negative real $s$ axis that arises from the analytic form of the incomplete Gamma function. Thus the FOPT Borel sum for the Adler function has the analytic property one expects from the physical Adler function. In contrast, the CIPT result in
Eq.(\ref{eq:IRBorelDCIPTsum}) exhibits the  Landau pole of the strong coupling and the cut along the positive real $s$-axis, but is also exhibits the uncancelled unphysical cut along the negative real $s$-axis coming from the incomplete Gamma function.

What does the cut in $\hat D_{\rm Borel}^{\rm CIPT}(s)$ along the negative real $s$-axis that arises from IR renormalons mean? First, we cannot simply dismiss $\hat D_{\rm Borel}^{\rm CIPT}(s)$ because it is contained in the Borel representation of the CIPT moment series.
Disregarding for an instant the unphysical character of this cut, one can immediately understand from Fig.~\ref{fig:ucontour} that the CIPT and FOPT Borel sums in Eqs.(\ref{eq:AdlerBorelCIPT}) and (\ref{eq:AdlerBorelFOPT}) correspond to two different prescriptions to make the Borel integrals well-defined in the presence of IR renormalons. The FOPT Borel sum is the average of paths~1a and 1b. In contrast, the CIPT Borel sum is the result of path~1a for ${\rm Im}[x]<0$ and of path~1b for ${\rm Im}[x]>0$ (because one can deform path~2 as long as no cuts or poles are crossed). The prescriptions used in $\hat D_{\rm Borel}^{\rm CIPT}(s)$ and  $\hat D_{\rm Borel}^{\rm FOPT}(s)$ are
essentially different implementations of an IR cutoff and thus entail a particular prescription of the OPE corrections such that the hadron level prediction (i.e.\ the sum of perturbative and OPE corrections) is prescription independent. Since the hadron level Adler function does not have the unphysical cut, we can draw the conclusion that the unphysical cut contained in the Borel sum $\hat D_{p,\gamma,{\rm Borel}}^{\rm CIPT}(s)$ is compensated by a corresponding unphysical cut in the Borel-resummed OPE corrections.\footnote{This refers to the OPE corrections at the level of using the Borel sum as the value of the perturbation series, as mentioned below Eq.(\ref{eq:BorelFOPT}).} These OPE corrections can therefore in general not be of the standard type described in Eq.(\ref{eq:AdlerOPE}). Since the results in Eqs.(\ref{eq:IRBorelDCIPTsum}) and (\ref{eq:IRBorelDFOPTsum}) apply also in full QCD, this statement is also true beyond the large-$\beta_0$ approximation. 

In the large-$\beta_0$ approximation the form of the non-standard contributions in the Borel-resummed CIPT OPE corrections for single and double IR renormalon poles can be visualized by evaluating the difference of Eqs.(\ref{eq:IRBorelDFOPTsum}) and (\ref{eq:IRBorelDCIPTsum}) for $\gamma=1,2$:
\begin{eqnarray}
\label{eq:nonanap1}
\hat D_{p,1,{\rm Borel}}^{\rm FOPT}(s)-\hat D_{p,1,{\rm Borel}}^{\rm CIPT}(s) & = &
- {\rm sig}[{\rm Im}[s]]\,(i\pi)\,\frac{\Lambda_{\rm QCD}^{2p}}{(-s)^{p}} \,,\\
\label{eq:nonanap2}
\hat D_{p,2,{\rm Borel}}^{\rm FOPT}(s)-\hat D_{p,2,{\rm Borel}}^{\rm CIPT}(s) & = &
- {\rm sig}[{\rm Im}[s]]\,(i\pi)\,(a(-x))^{-1}\,\frac{\Lambda_{\rm QCD}^{2p}}{(-s)^{p}}\,.
\end{eqnarray}
They match precisely with the generic form of the OPE shown in Eq.(\ref{eq:AdlerOPE}) apart from the non-analytic factor ${\rm sig}[{\rm Im}[s]](i\pi)$. The factor prevents that any single IR renormalon pole contribution
in $\hat D_{p,1,{\rm Borel}}^{\rm CIPT}(s)$ can vanish in the contour integration for the spectral function moments and also explains the observation we made for CIPT moment series with $p\neq m$ at the end of Sec.~\ref{sec:analytic}.
It is remarkable that the different character of the FOPT and CIPT series, which are both defined within dimensional regularization and the $\overline {\rm MS}$ renormalization, can lead to such an effect. 

So far we have motivated the CIPT Borel representation $\hat D_{\rm Borel}^{\rm CIPT}(s)$ for the Adler function from the form of the CIPT Borel representation of the spectral function moments in Eq.(\ref{eq:BorelCIPT}). At this point the obvious question to be asked is, if  $\hat D_{\rm Borel}^{\rm CIPT}(s)$ is also relevant for the perturbative expansion of the Adler function $\hat D(s)$ itself for complex $s$ away from the Euclidean axis. To this end let us consider the two expansion approaches in terms of the complex-valued $\alpha_s(-s)$ and in terms of the real-valued $\alpha_s(s_0)$:
\begin{eqnarray}
\label{eq:AdlerseriesCIPT2}
\hat D^{\rm CIPT}(s) & \, =  \,& \, \sum_{n=1}^\infty
\Big({\textstyle  \bar c_{n} \,(\frac{a(-x)}{a_0})^n}\Big)a_0^n \, = \, \sum_{n=1}^\infty
\textstyle  \bar c_{n} \,a^n(-x) \,, 
\\ \label{eq:AdlerseriesFOPT2}
\hat D^{\rm FOPT}(s) & \, = \, &
\sum\limits_{m=1}^\infty\,
a_0^m \, \sum\limits_{i=0}^{m-1} \frac{(m-1)!}{i!(m-i-1)!}\,\bar c_{m-i}\,(-\ln(-x))^{i}\,.
%\, \sum\limits_{n=1}^\infty\,
%\big({\textstyle \frac{\alpha_s(s_0)}{\pi}}\big)^n \, \sum\limits_{k=1}^{n} k\, c_{n,k}\,\ln^{k-1}({\textstyle \frac{-s}{s_0}}) \,.
\end{eqnarray}
We again call them CIPT and FOPT expansions just due to their association to the CIPT and FOPT spectral function moments. It is obvious that the FOPT expansion $\hat D^{\rm FOPT}(s)$ is associated to the Borel representation $\hat D_{\rm Borel}^{\rm FOPT}(s)$ in Eq.(\ref{eq:AdlerBorelFOPT}) since we can apply the proof of Sec.~\ref{sec:representations}.
For the CIPT expansion $\hat D^{\rm CIPT}(s)$, however, the association to the Borel representation $\hat D_{\rm Borel}^{\rm CIPT}(s)$ does not appear imperative since we cannot use the argument at the beginning of Sec.~\ref{sec:representations} that the contour integration enforces that powers of $\alpha_s(-s)$ should be considered as part of the series coefficients. Moreover, we recall that at the level of perturbation theory (i.e.\ when considering the Taylor expansion of $B[\hat D](u)$) the expressions for $\hat D_{\rm Borel}^{\rm CIPT}(s)$ and $\hat D_{\rm Borel}^{\rm FOPT}(s)$ are equivalent.
So let us have a closer look at the actual numerical behavior of the CIPT and FOPT series for the Adler function in Eqs.(\ref{eq:AdlerseriesCIPT2}) and (\ref{eq:AdlerseriesFOPT2}), respectively, in comparison with the CIPT and FOPT Borel sums in Eqs.(\ref{eq:AdlerBorelCIPT}) and (\ref{eq:AdlerBorelFOPT}).

\begin{figure} 
	\centering
	\subfloat[\label{fig:AdlerCIPTRe}]{\includegraphics[width= 0.49\textwidth]{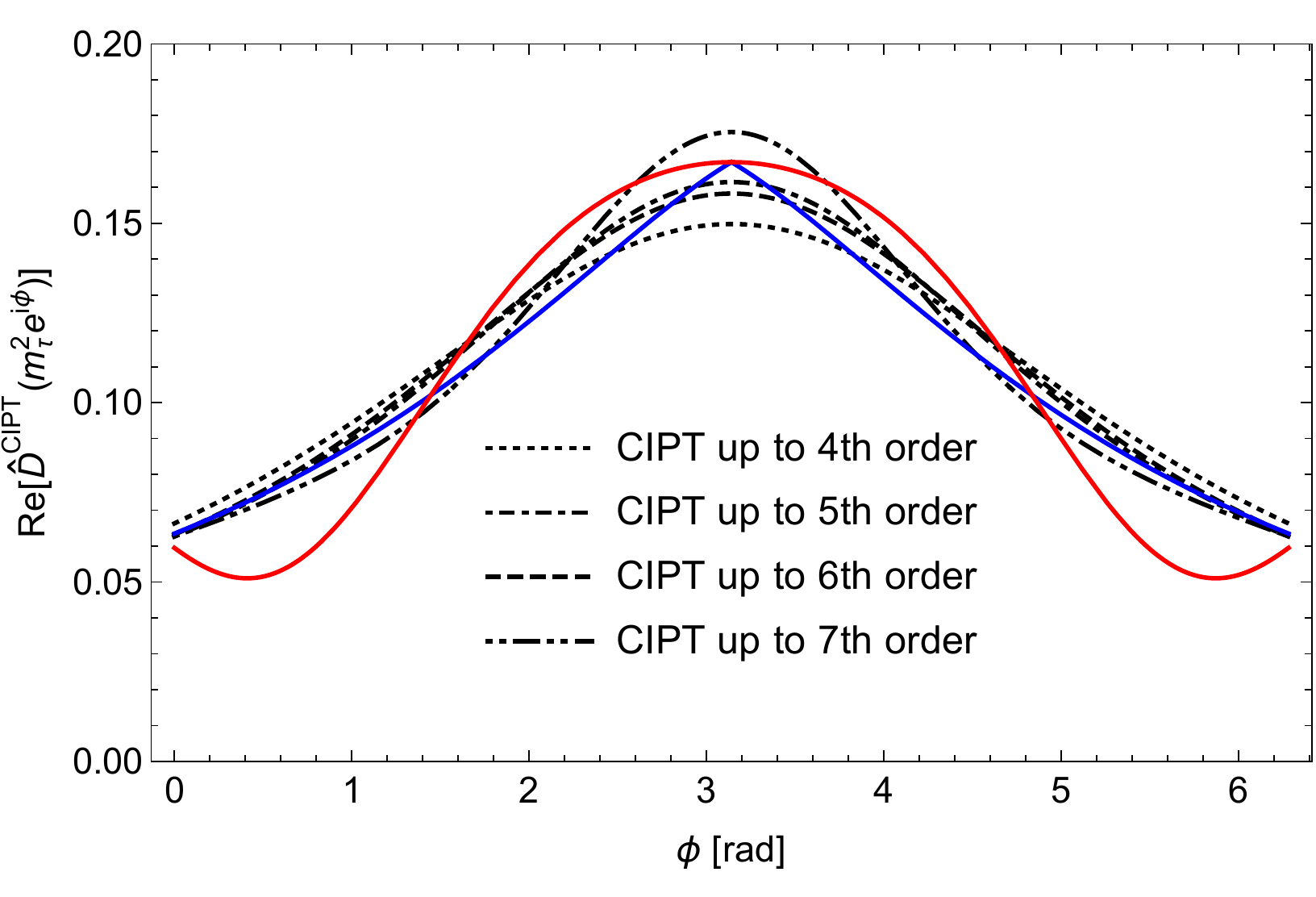}} ~~
%	\subfloat[\label{fig:AdlerCIPTIm}]{\includegraphics[width= 0.49\textwidth]{Adler_CIPT_Im_largeb0}}\\
	\subfloat[\label{fig:AdlerFOPTRe}]{\includegraphics[width=0.49\textwidth]{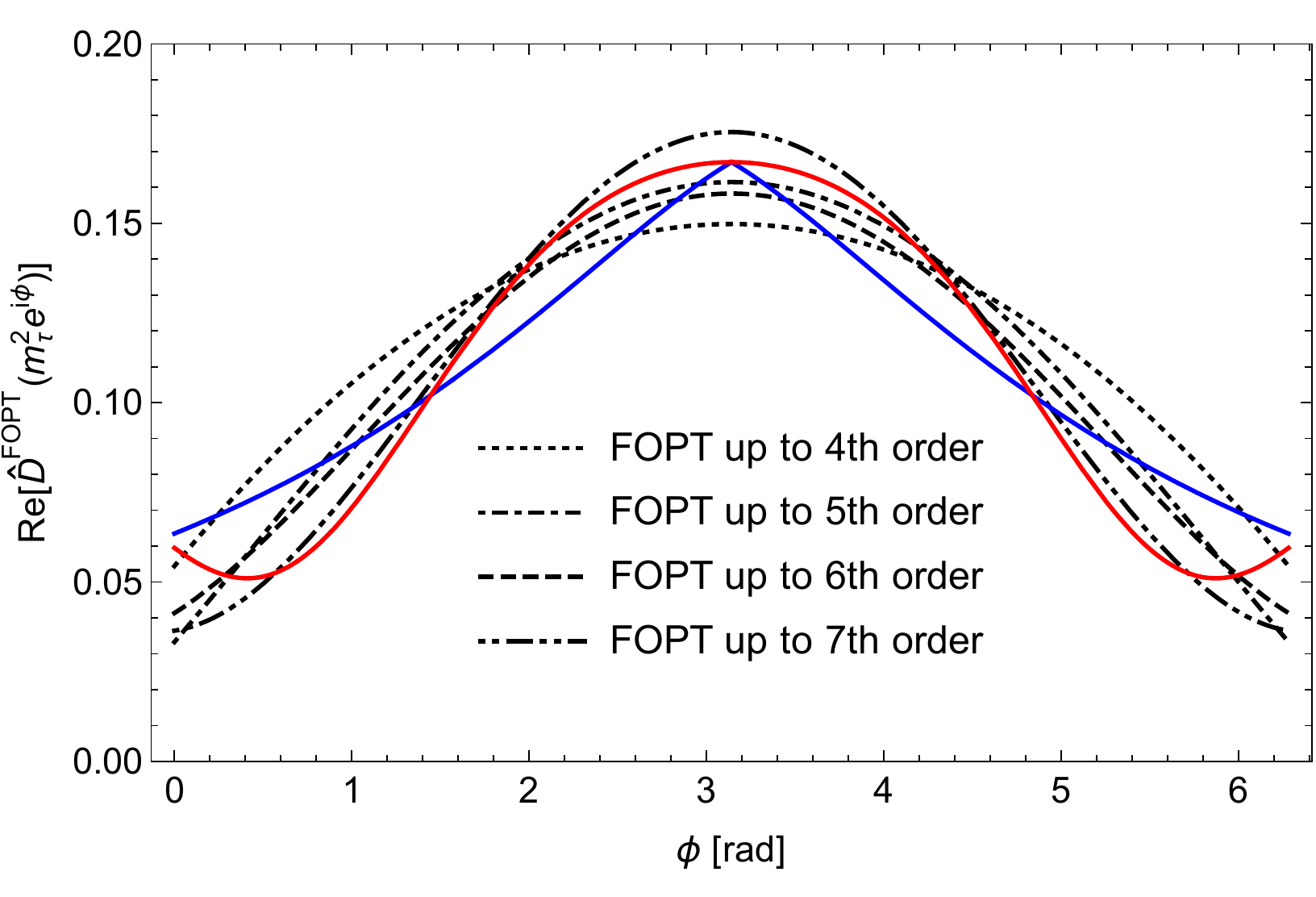}} %~~
%	\subfloat[\label{fig:AdlerFOPTIm}]{\includegraphics[width=0.49\textwidth]{Adler_FOPT_Im_largeb0}}
	\caption{\label{fig:AdlerfctBorel} 
		The real part of the Adler function series $\hat D^{\rm CIPT}(m_\tau^2 e^{i\phi})$ (panel a) and $\hat D^{\rm FOPT}(m_\tau^2 e^{i\phi})$ (panel b) for a summation of the perturbative series up to ${\cal O}(\alpha_s^4)$ (black dotted), ${\cal O}(\alpha_s^5)$ (black dot-dashed ), ${\cal O}(\alpha_s^6)$ (black dashed) and ${\cal O}(\alpha_s^7)$ (wide-dot-dashed) as a function of $\phi$ for $\alpha_s(m^2_\tau)=0.34$ in the large-$\beta_0$ approximation. Also displayed are the corresponding CIPT (blue) and FOPT (red) Borel sums.}
\end{figure}

In Fig.~\ref{fig:AdlerfctBorel} the real part of the Adler function $\hat D(s)$ for $s=m_{\tau}^2 e^{i \phi}$ is shown as a function of the angle $\phi$ in the large-$\beta_0$ approximation at the 4th (5-loop) to the 7th (8-loop) order (black lines) in CIPT (left panel) and FOPT (right panel) for $\alpha(m_\tau^2)=0.34$. In this range of orders the CIPT and FOPT both show a converging behavior.
We also displayed the CIPT (blue line) and FOPT (red line) Borel sums based on the generic expressions from Eqs.~(\ref{eq:UVBorelInt}), (\ref{eq:IRBorelDCIPTsum}) and (\ref{eq:IRBorelDFOPTsum}) for the Borel function in Eq.(\ref{eq:AdlerBorelb0}).
At the Euclidean point $\phi=\pi$ the FOPT and CIPT results agree as they must, but our focus is on the results at the other angles.
The CIPT series shows a well converging behavior, in particular close to the positive real axis (where $\phi$ is small or close to $2\pi$). But it is of course still an asymptotic series that eventually diverges. We clearly see that it approaches the CIPT Borel sum.
The FOPT series, on the other hand, is comparatively unstable close to the positive real $s$-axis, but it shows a reasonably good convergence behavior everywhere else. We see that it appears to approach the FOPT Borel sum and seems incompatible with the CIPT series values.
In Ref.~\cite{Hoang:2020mkw} the same analysis has been carried out in full QCD for the Borel model considered in Ref.~\cite{Beneke:2008ad} (which has a sizeable gluon condensate cut) and the outcome is the same.\footnote{The disparity between the FOPT and CIPT series for the real part of the Adler function was already noticed in Ref.~\cite{Beneke:2008ad}.} 

\begin{figure} 
	\centering
	\subfloat[\label{fig:ImAdlerCIPT}]{\includegraphics[width= 0.49\textwidth]{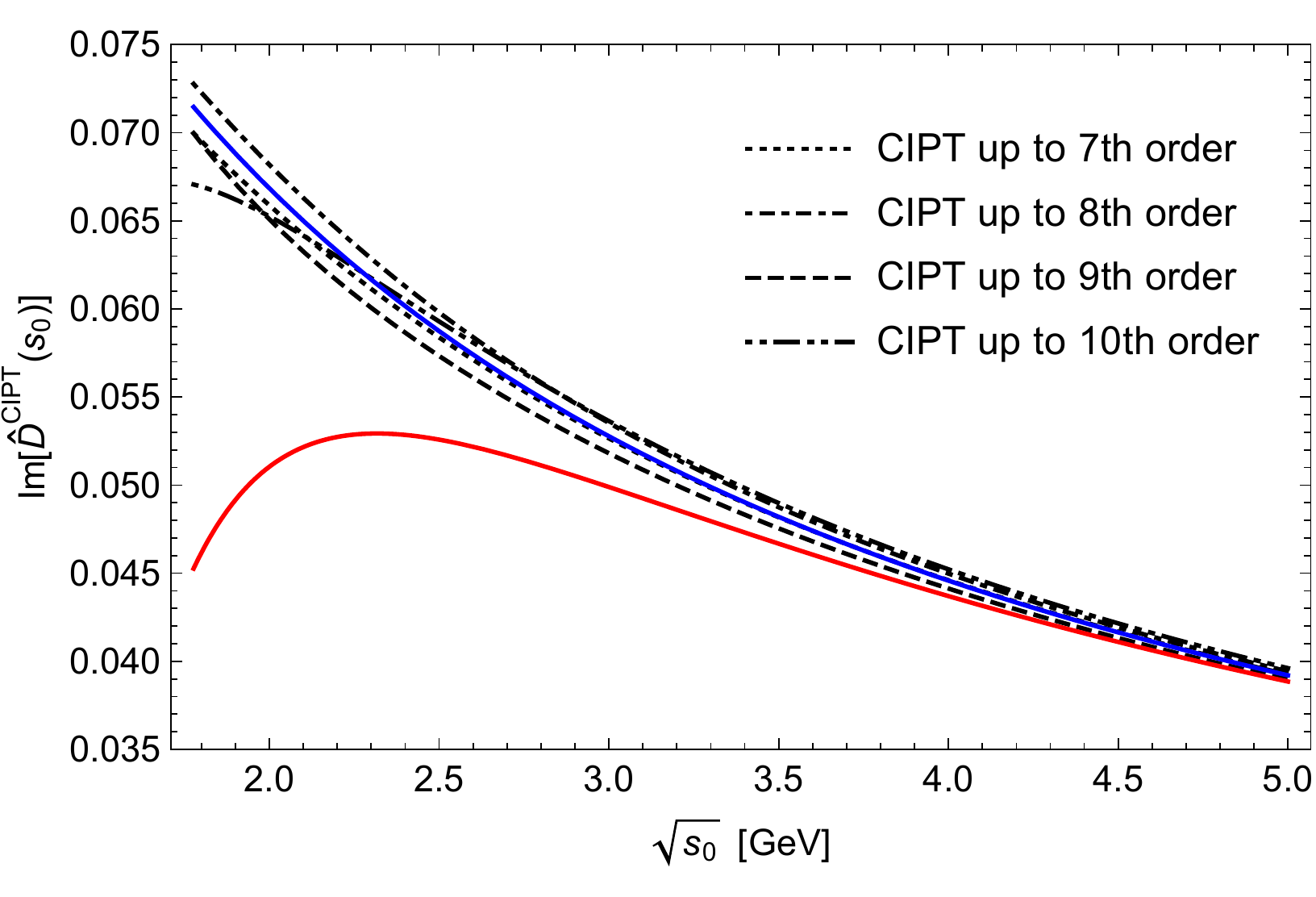}} ~~
	\subfloat[\label{fig:ImAdlerFOPT}]{\includegraphics[width=0.49\textwidth]{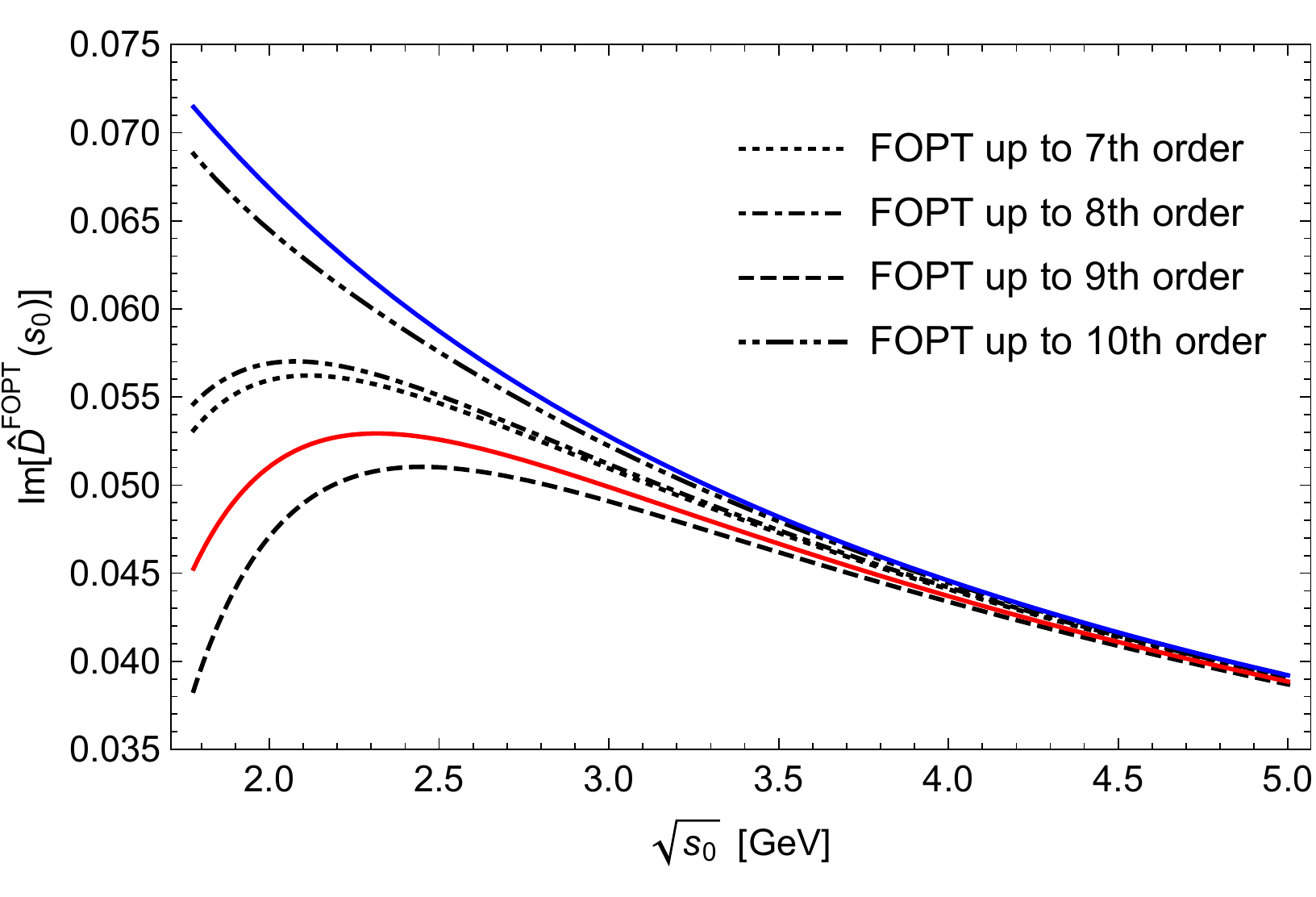}} 
	\caption{\label{fig:ImAdlerfctBorel} 
		Half of the discontinuity of the Adler function series along the real axis ${\rm Im}[\hat D^{\rm CIPT}(s_0+i0)]$ (panel a) and ${\rm Im}[\hat D^{\rm FOPT}(s_0+i0)]$ (panel b) for a summation of the perturbative series up to ${\cal O}(\alpha_s^7)$ (black dotted), ${\cal O}(\alpha_s^8)$ (black dot-dashed ), ${\cal O}(\alpha_s^9)$ (black dashed) and ${\cal O}(\alpha_s^{10})$ (black wide-dot-dashed) as a function of $\sqrt{s_0}$ for $\alpha_s(m^2_\tau)=0.34$ in the large-$\beta_0$ approximation. Also displayed are the corresponding CIPT (blue) and FOPT (red) Borel sums.}
\end{figure}

Another interesting quantity is the discontinuity of the Adler function along the positive real $s$ axis, which is related to the hadronic $R$-ratio. In Fig.~\ref{fig:ImAdlerfctBorel} the imaginary part of the Adler function ${\rm Im}[\hat D(s_0+i0)]$ is shown as a function of $\sqrt{s_0}$ in the large-$\beta_0$ approximation at the 7th (8-loop) to the 10th (11-loop) order (black lines) in CIPT (left panel) and FOPT (right panel). We again see that the CIPT series curves are quite stable and gather close to the CIPT Borel sum curve. The FOPT series, on the other hand, is comparatively unstable and scatters around the FOPT Borel sum curve. The $\sqrt{s_0}$-dependence of the difference between the CIPT and FOPT Borel sum curves visualized the $\Lambda_{\rm QCD}^4/s_0^2$ power dependence of the asymptotic separation. 

The essential observation we made in our analysis is, that the CIPT expansion of Eq.(\ref{eq:AdlerseriesCIPT2}), i.e.\ the expansion of the Adler function in terms of the complex-valued $\alpha_s(-s)$, where the physical cut is carried by the strong coupling, appears to be better described by the CIPT Borel representation  
$\hat D_{\rm Borel}^{\rm CIPT}(s)$ in Eq.(\ref{eq:AdlerBorelCIPT}), and not compatible with the FOPT Borel representation. 
The outcome in full QCD is the same for Borel function models with a sizeable gluon condensate cut. 
Of course, this interpretation relies on the view that the fact that the partial sums of the CIPT series at the orders where the corrections are minimal is compatible with the CIPT Borel sum is not accidental.

Overall, we found strong evidence that 
the Borel representation for the Adler function series in powers of $\alpha_s(-s)$ {\it away from the Euclidean axis} is the 
CIPT Borel representation in Eq.(\ref{eq:AdlerBorelCIPT}).
Thus, the non-standard character of the OPE corrections to the CIPT spectral function moments appears to already be a feature of the OPE corrections to the Adler function for the expansion in the complex-valued coupling $\alpha_s(-s)$ for non-Euclidean complex values $s$.

\section{Asymptotic Separation}
\label{sec:asymptoticseparation}

Let us now calculate and examine the asymptotic separation for the spectral function moments. Since the asymptotic separation arises only from the non-analytic IR renormalon contributions, it is sufficient to consider again the generic IR renormalon term in Eq.(\ref{eq:BorelDir}). In practice, in order to compute the CIPT Borel sum for a given Borel function (model), one identifies the IR singular terms of the form~(\ref{eq:BorelDir}), computes the asymptotic separation for each term and simply adds their sum to the FOPT Borel sum.  

The asymptotic separation arising from the generic IR renormalon term in Eq.(\ref{eq:BorelDir}) has the form
\begin{eqnarray}
\label{eq:Sepa1}
  \Delta(m,p,\gamma,s_0) \, \equiv \,
\delta_{\{(-x)^m,p,\gamma\},{\rm Borel}}^{(0),{\rm CIPT}}(s_0) \, - \,
\delta_{\{(-x)^m,p,\gamma\},{\rm Borel}}^{(0),{\rm FOPT}}(s_0) \nonumber \\[3mm]
 =\,
\frac{1}{2 \Gamma(\gamma)} \,\ointctrclockwise_{{\cal C}_x} \frac{{\rm d}x}{x} \, (-x)^m \,
{\rm sig}[{\rm Im}[x]]\,(a(-x))^{1-\gamma}\,e^{-\frac{p}{a(-x)}}
\,.
\end{eqnarray}
This result also applies in the context of full QCD.
As we already pointed out in Sec.~\ref{sec:analytic}, the path ${\cal C}_x$ starts/ends at $x_{\mp}=1\pm i0$. But due to the cut along the negative real $x$-axis, it is separated into two parts each of which always stay in the upper/lower imaginary half plane, such that the path never crosses the cut. The resulting integrations for $|x|\to\infty$ are non-trivial, since the suppression factor $e^{-\frac{p}{a(-x)}}\sim (\Lambda_{\rm QCD}^2/s_0)^p(-x)^{-p}$ competes with the enhancement factor $(-x)^m$. For the case $m<p$, where the exponential suppression wins, the upper path begins at $x_{-}=1+i0$ and proceeds to $x^\infty_-=-\infty+i\eta$ with $\eta$ being some positive real number. The lower path begins at $x^\infty_+=-\infty-i\eta$ and ends at $x_{+}=1-i0$. 
The cases $m=p$ and $m>p$ are more subtle, and one has to rely on an analytic continuation which is explained in more detail in Ref~\cite{Hoang:2020mkw}.
In the large-$\beta_0$ approximation the results for the asymptotic separation can be written down in a compact form  ($e^{-\frac{p}{a_0}}=\Lambda_{\rm QCD}^{2p}/s_0^{p}$),  
\begin{eqnarray}
\label{eq:Sepa4}
\Delta(p,p,\gamma,s_0)  & = & 0\,.\\ 
\label{eq:Sepabeta0} 
\Delta_{\beta_0}(m\neq p,p,1,s_0) & = &   \frac{ (-1)^{p-m}}{p-m}\,e^{-\frac{p}{a_0}} \,, 
\\
\Delta_{\beta_0}(m\neq p,p,2,s_0) & = &  (-1)^{p-m}\,\bigg[ \frac{1}{(p-m)^2} +  \frac{1}{(p-m)a_0}\bigg]\,e^{-\frac{p}{a_0}}\,, 
\end{eqnarray}
where the cases $m<p$ and $m>p$ are described by a single expressions related by analytic continuation. For the case $m=p$ the asymptotic separation vanishes. As expected, the results for the asymptotic separation inherit the power suppression $\sim \Lambda_{\rm QCD}^{2p}/s_0^{p}$ from the non-analytic terms shown in Eqs.(\ref{eq:nonanap1}) and (\ref{eq:nonanap2}).

It is highly instructive to compare the results for the asymptotic separation with the ambiguity that is commonly assigned to the FOPT Borel sum. The canonical definition of this ambiguity is obtained from the half of the difference obtained by doing the Borel integral along paths~1a and 1b times the conventional factor $i/\pi$:
\begin{eqnarray}
\label{eq:IRBorelIntFOPTambi}
 \frac{i}{2\pi} \left[\, 
\int_0^\infty \!\! {\rm d} u \,  \frac{e^{-\frac{u}{a(-x)}}}{(p+i 0 - u)^\gamma}
\, - \,
\int_0^\infty \!\! {\rm d} u \,  \frac{e^{-\frac{u}{a(-x)}}}{(p-i 0 - u)^\gamma}
\,\right] \, = \,
\frac{(a(-x))^{1-\gamma}}{\Gamma(\gamma)}\,e^{-\frac{p}{a(-x)}}\,.
\end{eqnarray}
This results in the following expression for the FOPT Borel sum ambiguity for the generic IR renormalon pole in  Eq.(\ref{eq:BorelDir}):
\begin{eqnarray}
\label{eq:IRBorelIntFOPTambi2}
\delta^{\rm FOPT}(m,p,\gamma,s_0) & \equiv &
\frac{1}{2 \pi i}\frac{1}{\Gamma(\gamma)} \,\ointctrclockwise_{|x|=1} \frac{{\rm d}x}{x} \, (-x)^m \,
(a(-x))^{1-\gamma}\,e^{-\frac{p}{a(-x)}}\,.
\end{eqnarray}
The expression is also valid in full QCD. In the large-$\beta_0$ approximation the results for the ambiguity can be readily calculated (see also Ref.~\cite{Beneke:2008ad}) and have the form
\begin{eqnarray}
\label{eq:ambitbeta01}
\delta^{\rm FOPT}_{\beta_0}(m\neq p,p,1,s_0) & = & 0 \,, \\ 
\label{eq:ambitbeta02}
\delta^{\rm FOPT}_{\beta_0}(p,p,1,s_0)  & = & e^{-\frac{p}{a_0}} \,, 
\\ 
\label{eq:ambitbeta03}
\delta^{\rm FOPT}_{\beta_0}(m\neq p,p,2,s_0) & = &  \frac{(-1)^{p-m}}{m-p}\,e^{-\frac{p}{a_0}}\,, 
\\ 
\label{eq:ambitbeta04}
\delta^{\rm FOPT}_{\beta_0}(p,p,2,s_0) & = & \frac{1}{a_0} \,e^{-\frac{p}{a_0}} \,. 
\end{eqnarray}
They exhibit the same kind of power-suppression as the asymptotic separation results, but their analytic form is quite different. 
It is also straightforward to see that for integer values $m\neq p$ we always have $\Delta > \delta^{\rm FOPT}$. 

However, the most conspicuous difference arises for a single pole ($\gamma=1$) IR renormalon if the weight functions do not contain the monomial $x^{p}$. This is the important case we have already mentioned in Sec.~\ref{sec:analytic} and which applies to the gluon condensate contributions for $p=2$. For Borel functions where the gluon condensate pole has a generic normalization (i.e.\ the pole's residue is not strongly suppressed), the asymptotic behavior arising from it dominates the Adler function series already at low orders.  
While the moment's FOPT Borel sum ambiguity in Eq.(\ref{eq:ambitbeta01}) vanishes identically,\footnote{In full QCD, as we mentioned already before, in this case the FOPT Borel sum ambiguity remains finite, but is strongly suppressed.} the asymptotic separation coming from the gluon condensate pole is finite due to Eq.(\ref{eq:Sepabeta0}) and directly proportional to $\Lambda_{\rm QCD}^4/s_0^2$. Parametrically this is of the same size as an unsuppressed standard-OPE gluon condensate correction would have, and therefore a very large effect. Since all phenomenologically relevant moments rely on the suppression of the gluon condensate in the FOPT approach, the asymptotic separation is therefore much larger than the FOPT Borel sum ambiguity, given that the Borel function contains a sizeable gluon condensate pole (or cut). For the large-$\beta_0$ approximation this is the case. 
But, as we have already mentioned in the introduction, in the literature there is no common agreement on whether the Adler function's Borel function  in full QCD has a sizeable gluon condensate cut (case A) or not (case B).

\begin{figure}
\centering
\subfloat[\label{fig:beta0simple0}]{\includegraphics[width= 0.49\textwidth]{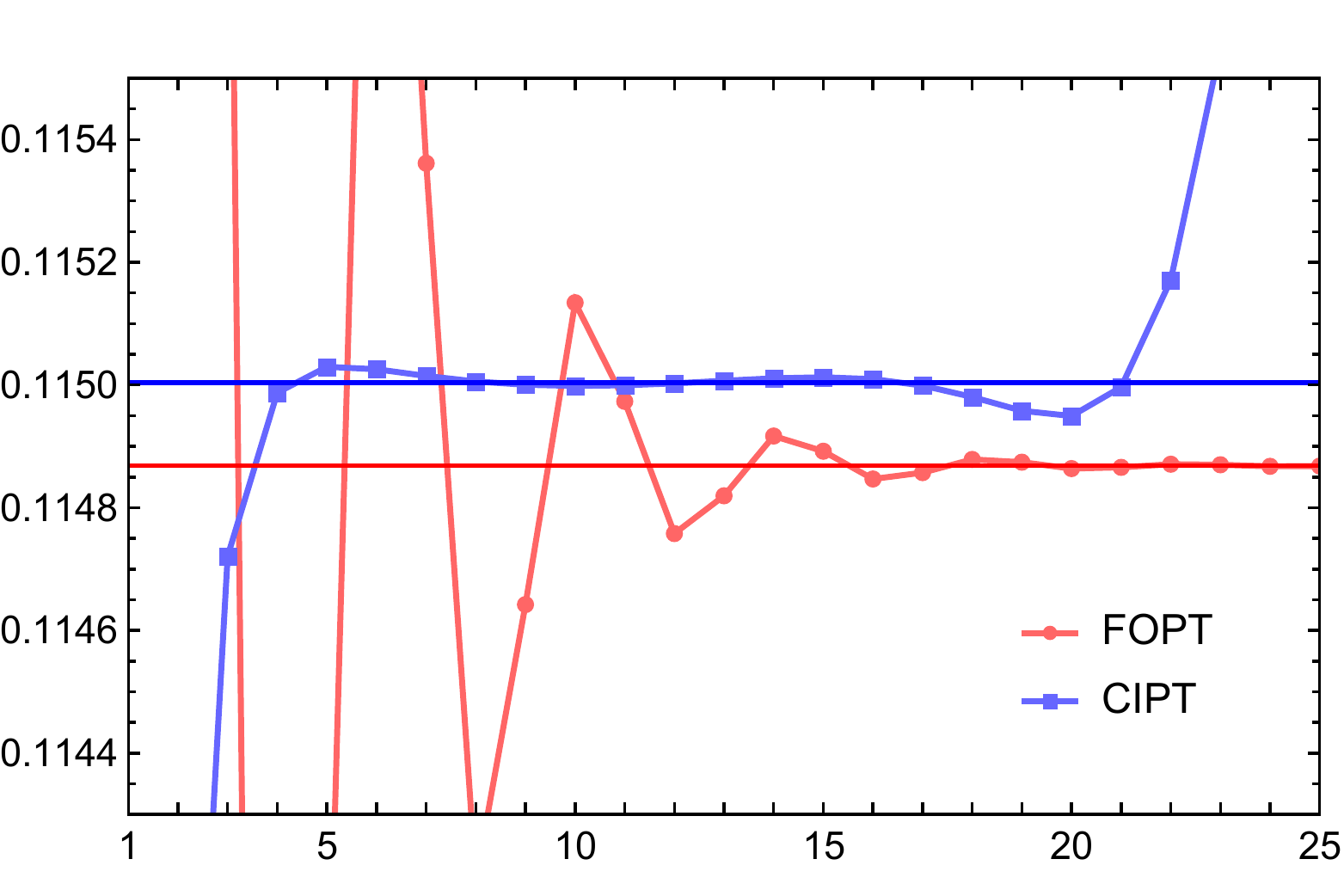}} ~~
%	\subfloat[\label{fig:beta0simple1}]{\includegraphics[width= 0.49\textwidth]{Simple_pole_m1}}\\
\subfloat[\label{fig:beta0simple2}]{\includegraphics[width=0.49\textwidth]{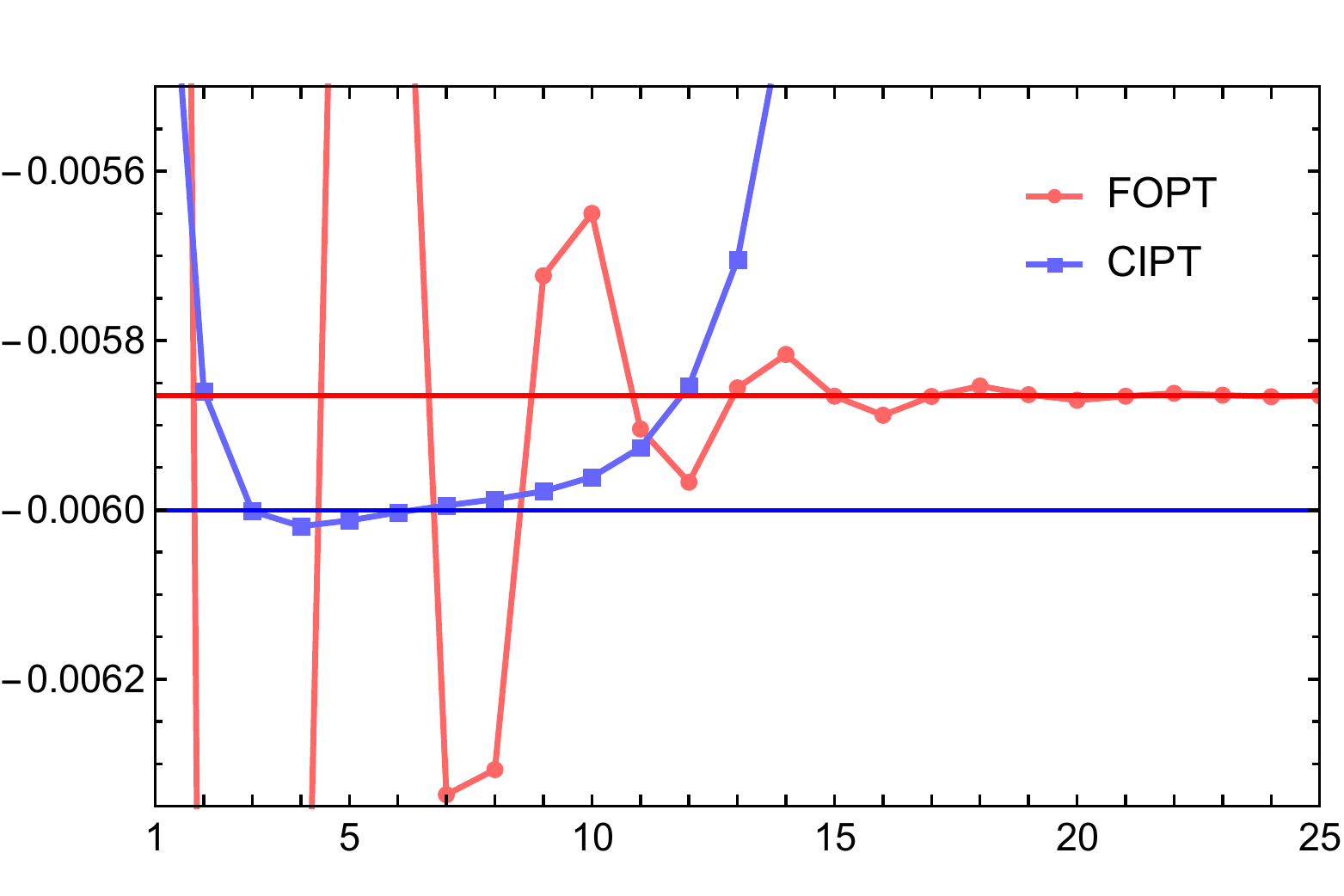}} %~~
%	\subfloat[\label{fig:beta0simple4}]{\includegraphics[width=0.49\textwidth]{Simple_pole_m4}}
\caption{\label{fig:beta0simplepole} 
	Moments $\delta_{\{(-x)^m,2,1\}}^{(0),{\rm FOPT}}(m_\tau^2)$ (red) and $\delta_{\{(-x)^m,2,1\}}^{(0),{\rm CIPT}}(m_\tau^2)$ (blue) in the large-$\beta_0$ approximation for a pure $p=2$ single renormalon pole and weight functions $W(x)=1$ (panel a) and  $W(x)=(-x)^4$ (panel b) as a function of the order up to ${\cal O}(\alpha_s^{25})$ in the large-$\beta_0$ approximation. The red and blue horizonal lines represent the FOPT and CIPT Borel sums, respectively.}
\end{figure}

Let us now examine the asymptotic separation for a number of concrete examples in the large-$\beta_0$ approximation. 
In Fig.~\ref{fig:beta0simplepole} we consider the spectral function moment series that is generated entirely from the Borel function  $B^{\rm IR}_{\hat D,2,1}(u)=1/(2-u)$, i.e.\  from a gluon condensate renormalon pole with residue one. It is instructive to study this case since the gluon condensate pole is responsible for the dominant contribution to the asymptotic separation. 
We consider the monomial moments with $W(x)=1$ (left panel) and $W(x)=(-x)^4$ (right panel). 
The FOPT and CIPT series are shown as the red and blue dots, respectively and we have connected subsequent orders by straight lines for better visualization. The red and blue horizonal lines are the FOPT and CIPT Borel sums. We recall that the FOPT series for $W(x)=(-x)^{m\neq 2}$ are convergent (i.e.\ $\delta^{\rm FOPT}_{\beta_0}(m\neq 2,2,1,m_\tau^2)=0$) and that the standard gluon condensate OPE correction vanishes. 
This means that the FOPT Borel sums represents an exact result without non-perturbative corrections. 
We see that FOPT series strongly oscillates initially, but eventually (mathematically) converges accurately to the FOPT Borel sum as expected. The CIPT series, on the other hand, is asymptotic and approaches a stable value at intermediate orders, but eventually diverges. The value that the CIPT series reaches at these intermediate orders, where the corrections are minimal, is quite stable and agrees very well with the CIPT Borel sum (which is the FOPT Borel sum plus the asymptotic separation $\Delta_{\beta_0}(m\neq 2,2,1,m_\tau^2)$). 
We furthermore observe that the range of order where the CIPT series is close to the CIPT Borel sum depends significantly on the monomial exponent. The outcome is the same for any weight function $W(x)=(-x)^{m\neq 2}$.
Since the FOPT Borel sum is the exact (hadron-level) result, the asymptotic separation quantifies the non-standard gluon condensate OPE correction that needs to be subtracted from the CIPT series to bring it into agreement with the FOPT series.

\begin{figure} 
	\centering
	\subfloat[\label{fig:beta0Rtau}]{\includegraphics[width= 0.49\textwidth]{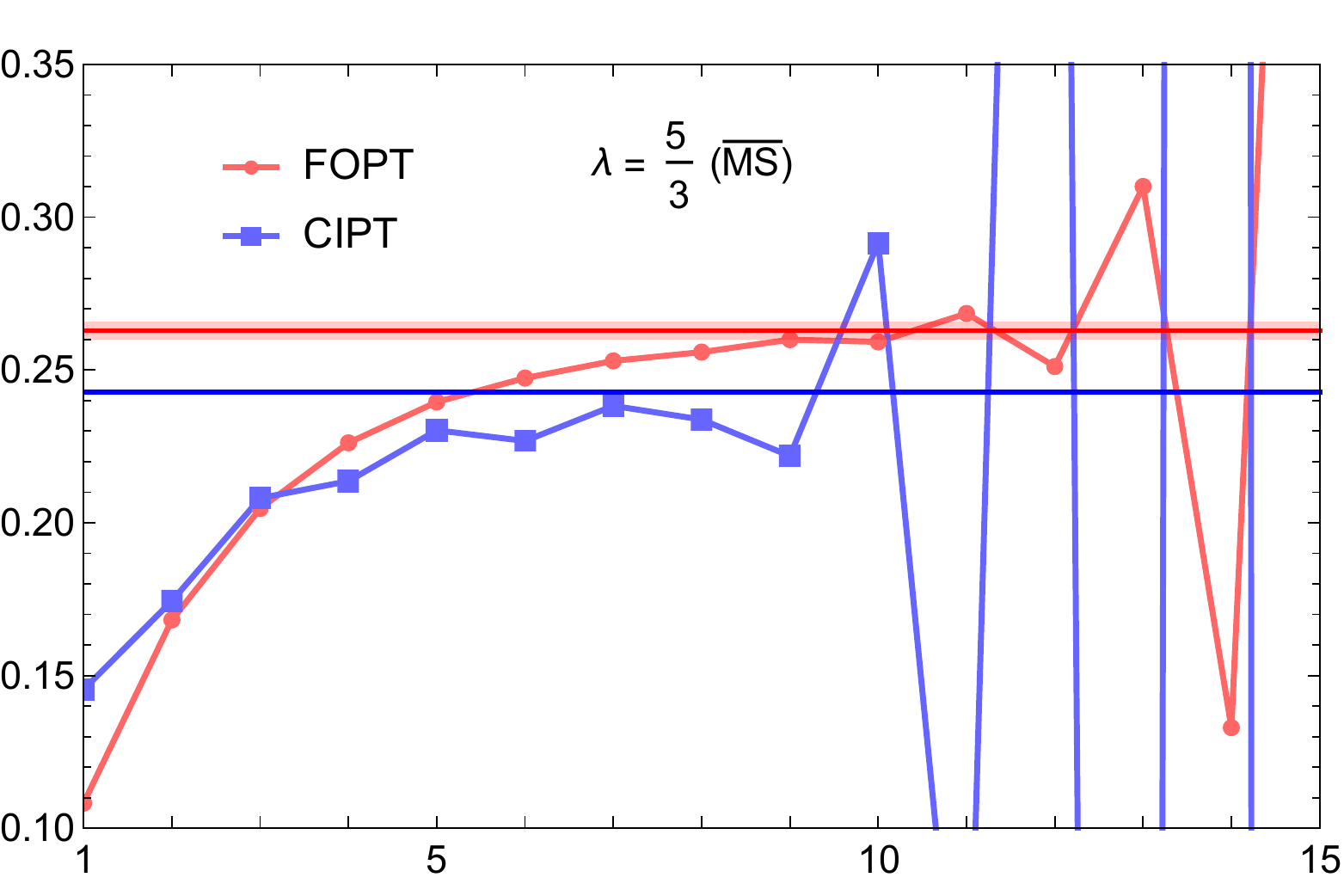}} ~~
	\subfloat[\label{fig:beta0Rtau5}]{\includegraphics[width=0.49\textwidth]{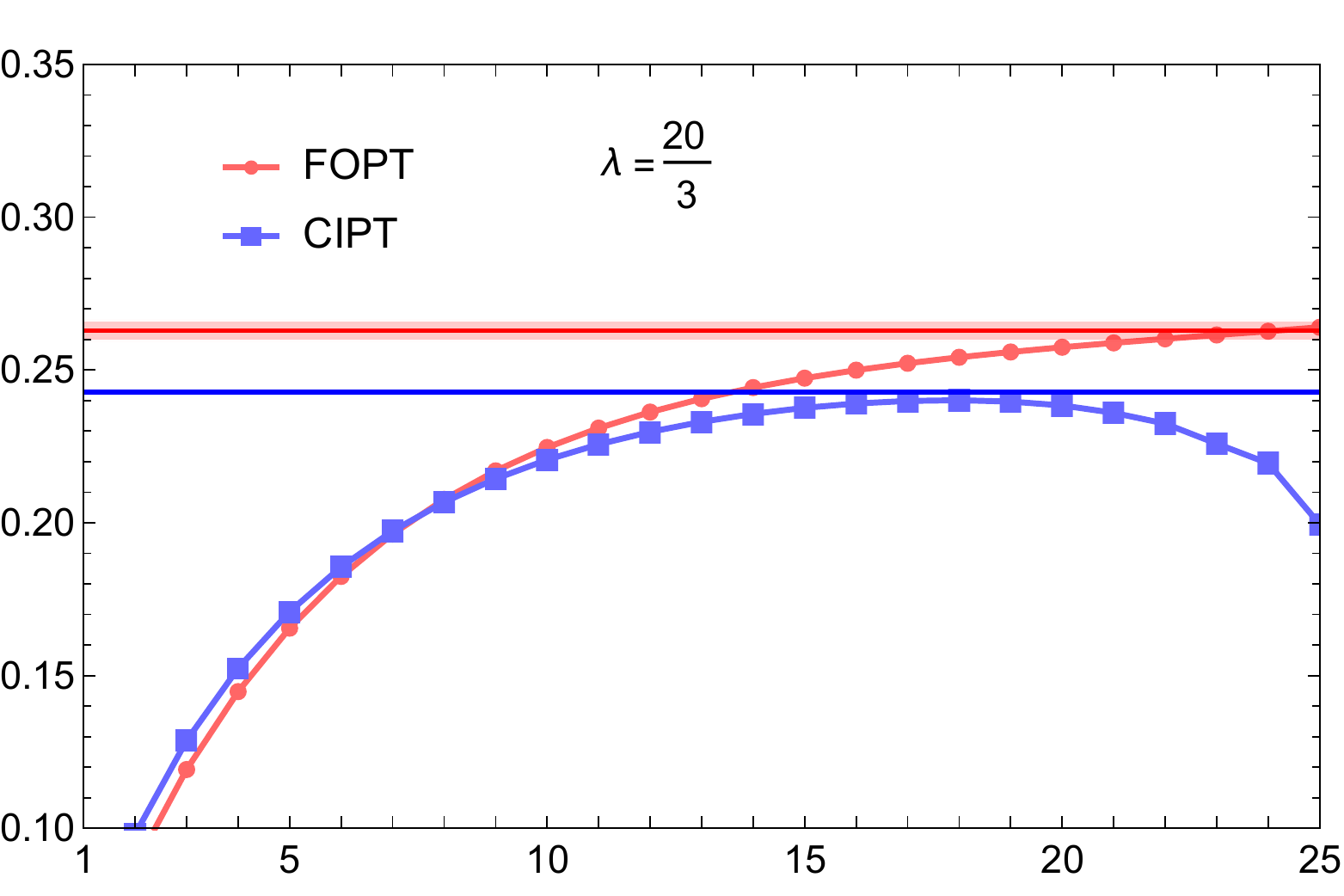}} %~~
	\caption{\label{fig:Rtau_beta0} 
		Moment series $\delta^{(0),{\rm FOPT}}_{W_\tau}(m_\tau^2)$ (red) and  $\delta^{(0),{\rm CIPT}}_{W_\tau}(m_\tau^2)$ (blue) in the large-$\beta_0$ approximation as a function of the truncation order. Panel (a) shows the result in the $\overline{\mbox{MS}}$ scheme and panel (b) in a scheme where the coupling is defined by $a^{(15/3)}(x) = a(x)/(1+3\times5/3\, a(x))$. 
		The red and blue horizonal lines represent the FOPT and CIPT Borel sums, respectively, and the orange bands indicate FOPT Borel sum ambiguity.}
\end{figure}

In Fig.~\ref{fig:Rtau_beta0} the FOPT and CIPT series for the hadronic tau decay width $R_\tau$ are shown in the large-$\beta_0$ approximation (i.e. using the all-order result for the Adler function~\cite{Broadhurst:1992si}, the Borel function in Eq.(\ref{eq:AdlerBorelb0}) and the kinematic weight function $W_\tau(x)=(1-2x+2x^3-x^4$)). The color coding is the same as for Fig.~\ref{fig:beta0simplepole}. For $R_\tau$ the gluon condensate renormalon is again eliminated and the gluon condensate standard OPE correction of Eq.(\ref{eq:deltaOPEdef}) vanishes. However, there are single and double IR renormalon poles for all integers $p\ge 3$, so that the FOPT series is asymptotic and the FOPT Borel sum ambiguity is finite. The FOPT Borel sum ambiguity is indicated by the horizontal light red band. Panel (a) displays the results in the $\overline{\rm MS}$ scheme. The FOPT and the CIPT series reflect the behavior known from full QCD (at the 5-loop level), namely that they show a well converging behavior, but that the CIPT series clearly approaches a smaller value. We again see that the FOPT series approaches the FOPT Borel sum, and the CIPT approaches the CIPT Borel sum (obtained from the sum of the FOPT Borel sum and the asymptotic separation). 
The asymptotic separation again quantifies the disparity of the two series very well. 
The agreement appears, however, not to be perfect due to the oscillatory behavior related to the UV renormalons. The impact of the UV renormalons can be suppressed using a different scheme for the strong coupling. An example is the geometric scheme change $a^{(15/3)}(x) \equiv a(x)/(1+3\times5/3 a(x))$ ($\alpha_s^{(15/3)}(m_\tau^2)=0.15332$), which leads to a strong suppression of the UV renormalon residues.\footnote{This scheme yields a strong coupling value that is much smaller than for $\overline{\rm MS}$ so that the series approach their Borel sums at much higher order. It is therefore not very useful for phenomenological applications.}   
The results in this scheme are shown in panel (b). We now see that the value of the CIPT series in the region where the corrections are minimal agrees very nicely with the CIPT Borel sum. As for the example discussed in Fig.~\ref{fig:beta0simplepole}, the non-standard contribution of the OPE corrections that would need to be added to the CIPT series exactly compensate the asymptotic separation. For $R_\tau$ it is worth to note that $99.8\%$ of the asymptotic separation originates from the gluon condensate $p=2$ pole in Eq.(\ref{eq:AdlerBorelb0}). This shows the strong numerical impact of the gluon condensate renormalon on the asymptotic separation and its implications.

We refer to Ref.~\cite{Hoang:2020mkw} for a more extended and detailed analysis. There we have also demonstrated that the analytic knowledge of the 
asymptotic separation can be used to design spectral function moments for which the asymptotic separation is much smaller than for $R_\tau$.
Overall, the numerical examinations demonstrate that the asymptotic separation correctly quantifies the discrepancy between the CIPT and FOPT series for spectral function moments in the order regions where the corrections are minimal. 

We reiterate, however, that concrete and definite quantitative statements can only be made if the Borel function, i.e. the normalization of the various non-analytic renormalon terms, is known. In the large-$\beta_0$ approximation this is the case. The reason why the asymptotic separation is much larger than the FOPT Borel sum ambiguity for $R_\tau$ is the sizeable residue of the gluon condensate $p=2$ renormalon pole in Eq.(\ref{eq:AdlerBorelb0}).
In full QCD the knowledge of the Adler function's Borel function is limited and - as we already pointed out before - there is no common consensus in the literature on the normalization of the gluon condensate cut. So, while the asymptotic separation and also the non-standard character of the OPE corrections to the CIPT series also apply in full QCD, their numerical impact is model-dependent.
If the normalization of the gluon condensate cut is sizeable, then the discrepancy between the CIPT and FOPT series at 5-loops is a systematic effect related to the gluon condensate cut and may be reconciled using the asymptotic separation and an associated non-standard OPE correction. If the normalization of the gluon condensate cut is strongly suppressed,\footnote{The normalization of the gluon condensate cut cannot vanish since the gluon condensate OPE correction is known to exist for the Adler function~\cite{Shifman:1978bx}.} the discrepancy between the CIPT and FOPT series at 5-loops is an unrelated issue and may be reconciled by higher-order corrections yet to be computed. For $\tau$ spectral function moment analyses within the CIPT expansion, the uncertainty concerning the correct form of the OPE corrections related to these two scenarios adds an additional source of theoretical error, which is absent when the FOPT expansion is used.

\section{Summary}
\label{sec:fullQCD}

In this article we have reported on a novel view on the discrepancy between the CIPT and FOPT approaches for the $\tau$ hadronic spectral function moments that was recently advocated in Ref.~\cite{Hoang:2020mkw}. Using the explicit order-by-order Borel transformations for the CIPT and FOPT spectral function moment series, two inequivalent Borel representations emerge due to the moment's contour integration which enforce the evaluation of the Adler function in the complex plane. The analytic properties of the two Borel representations are different because of IR renormalons which render the series asymptotic and which are tied to the known structure of the power corrections in the context of the OPE. While the FOPT Borel representation has been known before, the CIPT Borel representation is new and entails a particular regularization of the inverse Borel integration that differs from the common PV prescription used for FOPT. As a result, the Borel sums of the CIPT and FOPT series do not agree. Their difference, called the 'asymptotic separation', can be calculated analytically. The asymptotic separation is significantly larger than the ambiguity that is commonly assigned to the FOPT Borel sum if the dimension-4 gluon condensate renormalon cut in the Adler function's Borel function has a sizeable normalization.  
Using all-order results for the Adler function in the large-$\beta_0$ approximation (and concrete Borel function models in full QCD in Ref.~\cite{Hoang:2020mkw}) we have shown that the discrepancy in the asymptotic high-order behavior for the CIPT and FOPT moment series is indeed quantified by the asymptotic separation. 

An important implication of the existence of the asymptotic separation is that the OPE corrections that need to be added to the CIPT and FOPT moments differ. The analytic structure of the CIPT Borel representation further implies that the CIPT OPE corrections do not have the standard form that has been assumed in previous phenomenological analyses. In the large-$\beta_0$ approximation, where the Adler function's Borel function is known exactly, these implications are imperative. 
We have furthermore found strong evidence that the asymptotic separation already exists at the level of the Adler function away from the Euclidean region, where the expansion in the complex-valued coupling $\alpha_s(-s)$ corresponds to the CIPT Borel representation.  

At this time, there is no common agreement in the literature on whether the Borel function of the Adler function in full QCD has a sizeable gluon condensate singularity or not. If it does, our findings may provide an explanation for the discrepancy that is observed for the CIPT and FOPT spectral function moments which are currently known at the 5-loop level. 
The asymptotic separation may then allow to reconcile the CIPT and FOPT approaches. If it does not, the observed 5-loop discrepancy is an unrelated issue. The findings of our work imply that for $\tau$ spectral function moment analyses within the CIPT expansion, the uncertainty concerning the correct form of the OPE corrections related to these two scenarios adds an additional source of theoretical error, which is absent when the FOPT expansion is used.

\section*{Acknowledgments}
This work was supported in part by the FWF Austrian Science Fund under the Project No. P32383-N27.
We acknowledge partial support by the FWF Austrian Science Fund under the Doctoral Program ``Particles and Interactions'' No.\ W1252-N27. We also thank the Erwin-Schr\"odinger International Institute for Mathematics and Physics for partial support.

\end{document}